\documentclass{sig-alternate}
\usepackage{epsfig,endnotes,comment,hyperref,amssymb,amsmath,array}
\usepackage{multirow}
\usepackage{flushend}
\usepackage{hyperref}

\newcommand{\code}[1]{{\small \tt #1}}
\makeatletter
\def\@copyrightspace{\relax}
\makeatother
\begin{document}

\date{}

\title{Deanonymisation of clients in Bitcoin P2P network}

\numberofauthors{3} 
%
\author{
%
%
\alignauthor
Alex Biryukov\\
       \affaddr{University of Luxembourg}\\
       \email{{\small alex.biryukov@uni.lu}}
\alignauthor
Dmitry Khovratovich\\
       \affaddr{University of Luxembourg}\\
       \email{{\small dmitry.khovratovich@uni.lu}}\\
\alignauthor
Ivan Pustogarov\\
       \affaddr{University of Luxembourg}\\
       \email{{\small ivan.pustogarov@uni.lu}}
}

\maketitle


\subsection*{Abstract}

Bitcoin is a digital currency which relies on a distributed set of miners to mint coins and on a 
peer-to-peer network to broadcast transactions.  The identities of Bitcoin users are hidden behind 
pseudonyms (public keys) which are recommended to be changed frequently in order to increase transaction unlinkability.

We present an efficient method to deanonymize Bitcoin users, which allows to link user pseudonyms to the IP addresses where the transactions are generated. 
Our techniques work for the most common and the most challenging scenario when users are behind NATs or firewalls of their ISPs.  They allow to link 
transactions of a user behind a NAT and to distinguish connections and transactions of different users behind the
same NAT. We also show that a natural countermeasure of using Tor or other anonymity services can be cut-off by abusing anti-DoS countermeasures of the Bitcoin network.
Our attacks require only a few machines and have been experimentally verified. We propose several countermeasures to mitigate these new attacks\footnote{A simplified description of the attack can be found \href{https://www.cryptolux.org/index.php/Bitcoin}{here}.}.


\section{INTRODUCTION}
\label{sec:intro}
Digital currency based on cryptography is not a new idea~\cite{Chaum} but
till recently it did not attract much attention. It changed rapidly
with introduction of Bitcoin~\cite{nakamoto2012bitcoin}. Bitcoin is a
decentralized digital currency which does not rely on a trusted issuing
entity but rather on a peer-to-peer network with peers minting Bitcoins by
brute-forcing double SHA-256 hash function. To make the money generation
process computationally hard,
the Bitcoin protocol requires the minters to present the hash value of a data block with new
portion of Bitcoins  and new transactions to have a certain number of
zeros (an instance of  the \emph{Proof-of-Work} concept).

Bitcoin is now accepted as a currency by many companies from online
retailer Overstock to exotic Virgin Galactic~\cite{bcvirgin}.
One of its main advantages over bank transfers is it's decentralized
architecture and absence of intermediaries. 
This prevents shutting it down or seizing by a government.
Bitcoin money transfers are
non-refundable, reasonably fast\footnote{The network sees a transaction immediately, but the receiver has to wait for 1-2 hours to be sure that there is no double-spending.} and allow to send
money to any part of the world. The Bitcoin peer network
consists of homogeneous nodes and provides peer discovery and reputation
mechanisms to achieve stability.
The number of Bitcoin peers is estimated to be about 100,000 nowadays. The vast majority of these peers (we call them \emph{clients}), about 90\%, are located behind NAT and
do not allow any incoming connections, whereas they choose 8 outgoing connections to \emph{servers} (Bitcoin peers with public IP).

In a Bitcoin transaction, the address of money sender(s) or receiver(s) is a hash of his public key. We call such address a \emph{pseudonym} to avoid confusion with the IP address of the host where transactions are generated, and the latter will be called just \emph{address} throughout the text. 
In the current Bitcoin protocol the entire transaction history is publicly
available so anyone can see how Bitcoins travel from one pseudonym to
another and potentially link different pseudonyms
of the same user together. A theoretical possibility
of such attack was already mentioned in the original Bitcoin 
paper~\cite{nakamoto2012bitcoin}. Since then several papers ~\cite{Meiklejohn,ron2013quantitative} showed that
it is indeed possible by analysing the  transaction graph to cluster pseudonyms to different users.
Combined with some other sources (e.g. forum posts), the
clusters (and thus the users) can sometimes be mapped to real
identities~\cite{reid2013analysis,Meiklejohn}.
Even so, these methods are not generic, 
and the problem of how to tie a Bitcoin address to an actual identity
remained unsolved.

Evidently, studying the entire IP traffic of the Bitcoin peers would reveal the origins of each transaction and disclose the identities of many users,
but how much can be achieved by an ordinary attacker with a few machines and no access to clients behind NAT has been unclear.

Lerner~\cite{lerner} and Koshy et al.~\cite{koshyanalysis} were the first who
attempted an attack in this direction.
A vulnerability which allowed to link IP addresses of clients to their
\code{bitciond} wallets' addresses was reported by Lerner~\cite{lerner}.
The vulnerability exploited a protection against ``penny-flooding''
which prevents a transaction with very low fees and big size to be
forwarded or even stored by a Bitcoin peer.
The protection tested if a transaction was from a wallet owned by the user,
and if it was the case, then the protection was skipped. This allowed an attacker
to test if a peer possessed a Bitcoin address
by sending him specifically crafted transactions. The vulnerability required that
the attacker had a connection to a peer (thus targeting either Bitcoin servers or clients
which established connections to the attacker).
This vulnerability was fixed since version 0.7.2.

Koshy et al.~\cite{koshyanalysis} managed to 
deanonymize 1162 addresses over the period of 5 months.
Their approach, however, is limited to the transactions that expose anomalous behaviour like transactions
relayed only once or transaction that were relayed multiple times by the
same IP. Secondly, the proposed method only allows to get IP addresses of servers, which constitute only 10\% of the network, and not of the clients.
Finally, their paper does not discuss the case when a Bitcoin peer protects
himself by proxying his transactions through the Tor anonymity network.

\paragraph{Our contributions}
In this paper we describe a generic method to deanonymize a significant fraction of Bitcoin users and correlate their pseudonyms with public IP addresses. The method explicitly targets the clients (i.e. peers behind NAT or firewalls) and can differentiate the nodes with the same public IP.  Furthermore, our method also handles the case when the clients use anonymity services like Tor.
If a client uses two different pseudonyms during a single session, and even if they are unrelated in the transaction graph so that the linkage would be totally unachievable via
the transaction graph analysis~\cite{Meiklejohn}, our method is likely to catch it and glue the pseudonyms together. The method is generic and might be used in other P2P networks.

The crucial idea is that each client can be uniquely identified by a set of nodes
he connects to (\emph{entry nodes}). We show that this set can be learned at the time of connection and then used to identify the origin of a transaction.

Our attack requires only a few machines that establish a certain number of connections by Bitcoin protocol and log the incoming traffic.
In a concrete example, an attacker with a few GB of storage and no more than 50 connections to each Bitcoin server can disclose the sender's IP address in 11\%
\footnote{The actual fraction of
deanonymized transactions depends on
how stealthy an attacker wants to be.} of all transactions generated in the Bitcoin
network. If the attacker allows a slight DoS of the network, he may achieve deanonymization rates up to 60\%, which has been confirmed by the experiments in the Bitcoin test network.
We estimate the cost of the attack on the full Bitcoin network to be under 1500 EUR per month.

The computational power needed to disclose the sender of a single transaction is negligible and is far smaller than the amount of work needed
to process the transaction graph in \cite{reid2013analysis,Meiklejohn}.
For the best of our knowledge this is the first attack which targets
Bitcoin peers behind NAT. Our attack does not assume any anomaly in the behaviour of peers or in the traffic
and would work even if Bitcoin would encrypt the connection. It might be applicable to other digital currencies derived from Bitcoin.

As another interesting though unrelated to deanonymisation idea we 
look at how to decrease block mining difficulty by creating
an alternative blockchain reality. This becomes important since Bitcoin
by design is not adaptive to rapid drops in hash power of miners
and might become necessary in case of many miners quit mining.
This is not just a hypothetical case, since Bitcoin exchange
rate can fall suddenly and rapidly, making block mining unprofitable.

\paragraph{Roadmap}
Our paper is structured as follows:
\begin{itemize}
 \item We give necessary background of how Bitcoin works and the rules its peers follow to broadcast their addresses and transactions.
 \item As a first step towards deanonymization, we show  how to prohibit Bitcoin clients from using the Tor
       anonymity network  by exploiting Bitcoin anti-DoS protection mechanism (Section~\ref{subsec:tor}).
 \item We show how to learn the connections of the Bitcoin clients in Section~\ref{sec:connectivity}.
 \item We finally show how to identify the sender of a transaction (i.e. deanonymize him) in Section~\ref{sec:deanon}. We recover the public IP address of the sender and 
further differentiate clients sharing
the same public IP. 
\item We discuss how to choose parameters of the attack and its success rate and explain our experiments on the test network. We also propose countermeasures to mitigate the attack.
  \item As an extra result, we outline a strategy to lower the difficulty of the system by adding a
        properly selected value to the list of checkpoints nodes hard-coded in the
        client code. It can be used by the entire community if the mining becomes
        unbearable and non-profitable, or by malicious administrators who want
        to ruin the system (Section~\ref{sec:alt-real}).
\end{itemize}

{\it Ethical considerations}. 
All vulnerabilities described in this paper were reported
to Bitcoin core developers. When possible we carried out
experiments in the Bitcoin test network. To protect user privacy, we restricted from performing a full-scale deanonymization in the real network.
However, gathering some
statistics required us conducting experiments
on the main network, which did not cause disruption or exposure of
the main network. 


\section{BACKGROUND}
\label{sec:background}

In this section we provide a basic overview of the Bitcoin system.
Originating from a proof-of-concept source code and an accompanying
it white paper~\cite{nakamoto2012bitcoin}, the Bitcoin protocol never
had been fully documented, and is de-facto
the functionality of the primary Bitcoin client,
\textsf{bitcoind}~\cite{bitcoind}. In the following text we provide
only the details of the protocol that are crucial to our research.
These details are accumulated from the source code of \textsf{bitcoind}
and to a large extent are explained in the informal wiki
documentation~\cite{bitcoin-wiki}.

\paragraph{Block chain}
Bitcoin operates on a list of blocks, the \emph{block chain}.
Each block contains a header and transaction
data\footnote{All these conditions are strictly enforced,
and a block not conforming to them is discarded immediately.}.
The 80-byte header $\mathrm{Head}$ contains the 256-bit hash of the
previous block $H_{i-1}$, the timestamp (in seconds) $T_i$, the 32-bit
nonce $N_i$ (used to generate blocks), the hash $TX_i$ of the
transaction data , and the difficulty parameter $d_i$. To be valid, the
double-hash of the block header must be smaller (as an integer)
than a certain value, which is  a linear function of the
difficulty parameter:
$$
H_i=\text{SHA-256}(\text{SHA-256}(H_{i-1}||T_i||TX_i||d_i||N_i||))) < f(d_i).
$$ 
Currently it must be smaller than $2^{192}$, i.e. have its 64 most
significant bits equal to zero.

The Bitcoin miners first collect all transactions not yet included into a block. Then they generate the header fields and exhaustively try
different nonces, timestamps, and other parameters in order to obtain a
valid block. They are rewarded by 25 BTC (about \$14,000 by current
market rate), which is the very first transaction in the attached
transaction list. 
Whenever a block is created, a miner broadcasts it to the network, so
that each node attaches it into its internal block chain.

Payers and payees of the system are identified in the blockchain by
their Bitcoin addresses, or {\it pseudonyms}. A pseudonym is the base58-encoding
of the hash of the corresponding public key.
Whenever a payer wants to transfer his coins to another user,
he generates a
transaction and signs it with his private key. Signed transactions are
then added to the blockchain by miners. By checking the signature,
other Bitcoin participants can verify the new ownership of the coins.

\subsection*{Bitcoin P2P network}
Peers of the Bitcoin network connect to each other over an unencrypted
TCP channel. There is no authentication functionality in the network,
so each node just keeps a list of IP addresses associated with its
connections. 

To avoid denial-of-service attacks, the Bitcoin protocol minimizes
the amount of information forwarded by peers. Valid blocks and transactions
are relayed whereas invalid blocks
are discarded. Moreover, Bitcoin implements a reputation-based protocol
with each node keeping a penalty score for every connection. 
Whenever a malformed message is sent to the node, the 
latter increases the penalty score of the connection and bans the
``misbehaving'' IP address for 24 hours when the penalty reaches the
value of 100.

Though official Bitcoind software does not explicitly divide its
functionality between clients and servers, Bitcoin peers can be grouped
into those which can accept incoming connections (servers) and
those which can't (clients), i.e. peers behind NAT or firewall, etc.
At the time of writing there were about 8,000 reachable servers while
the estimated number of clients was about 100,000. 

\begin{figure}[h]
\begin{center}
\includegraphics[scale=0.3]{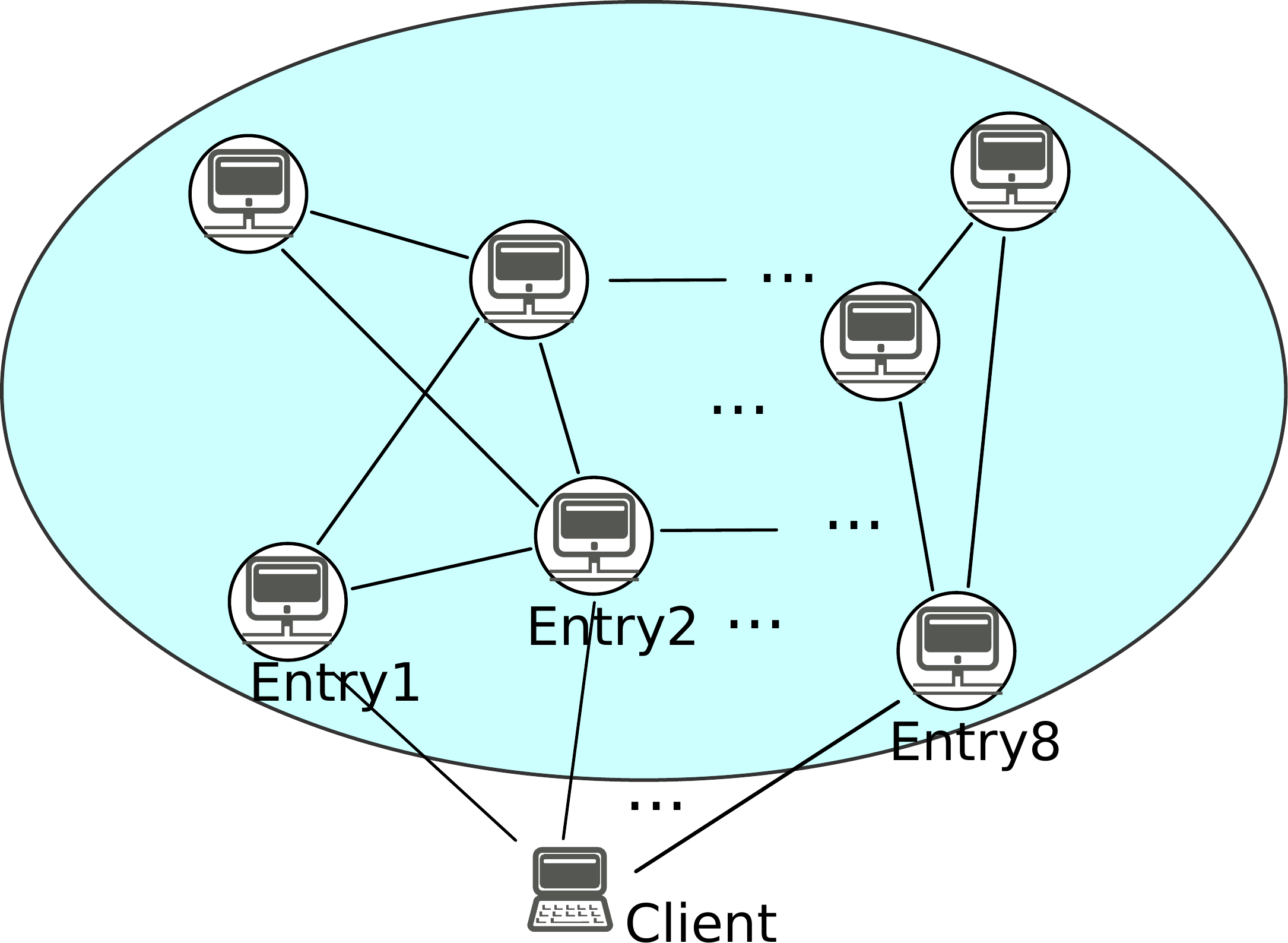}
\caption{Bitcoin network}
\label{fig:architecture}
\end{center}
\end{figure}

By default Bitcoin peers (both clients and servers) try to maintain
8 outgoing connections. In addition, Bitcoin servers can accept up
to 117 incoming connections (thus having up to 125
connections in total). If any of the 8 outgoing connections drop, a
Bitcoin peer tries to replace them with new connections. If none of
the 8 outgoing connections drop, the peer will stay connected to them
until it is restarted. In case of a client, we call the 8 nodes to
which it establishes connections {\it entry nodes}
(see Fig.~\ref{fig:architecture}). A Bitcoin server accepts
any number of connections from a single IP address
as long as the treshold for the total number of connections is not reached.

\paragraph{Address propagation}
The Bitcoin
protocol implements an address propagation mechanism to help peers to discover other peers in the P2P network.
Each Bitcoin peer maintains a list of addresses of other peers in
the network and each address is given a timestamp which determines
its freshness.
Peers can request addresses from this list from each other
using \code{GETADDR} messages and unsolicitely advertise
addresses known to them using \code{ADDR} messages\footnote{One
\code{ADDR} message can contain any number of address, however messages
containing more than 1000 addresses are rejected on the remote side.}.
Whenever a Bitcoin node receives an \code{ADDR} message it decides
individually for each address in the message if to forward it to its neighbours.
It first checks if (1) the total number of addresses in the
corresponding \code{ADDR} message does not exceed 10, and (2) the attached timestamp 
is no older than 10 minutes. If either of these two checks fails, the
address is not forwarded; otherwise the address is scheduled for
forwarding\footnote{By scheduling a transmission we mean that the
node puts the corresponding message to the outgoing queue but does not yet
make the actual transmission.} to two of the node's neighbours in case the
address is reachable and to one neighbour only if it is non-reachable.
An address is considered {\it reachable} by a node if the node has a
network interface associated with same address family. Otherwise the
address is marked as {\it unreachable}.
According to the current reference implementation Bitcoin nodes recognize
three types of addresses:
IPv4, IPv6, and OnionCat addresses~\cite{onioncat}. 
Limiting the number of
neighbours to which an address is forwarded reduces the total amount of
traffic in the Bitcoin P2P network.

In order to choose neighbours to which to forward an address, a Bitcoin
node does the following. 
For each of its neighbours it computes a hash of a value composed of
the following items: address to be forwarded, a secret salt, current
day, and the memory address of the data structure describing the
neighbour. The exact expression for the hashed value is of little
importance for our attacks. The only thing which we need to emphasize is
that the hash stays the same for 24 hours.
The peer then sorts the list of its neighbours based on the
computed hashes and chooses the first entry or two first entries (which
depends on the reachability of the address). In the rest of the paper
we call such nodes {\it responsible nodes} for the address.

The actual transmission of the scheduled \code{ADDR} messages does not happen
immediately. Every 100 milliseconds one neighbour is randomly selected from the
list off all peer's neighbours and the queue for outgoing \code{ADDR}
messages is flushed for this node only.
We call the node chosen at the beginning of a 100 milliseconds round 
{\it trickle node} and the procedure as a whole as {\it trickling}.

\begin{figure}[h]
\begin{center}
\includegraphics[scale=0.3]{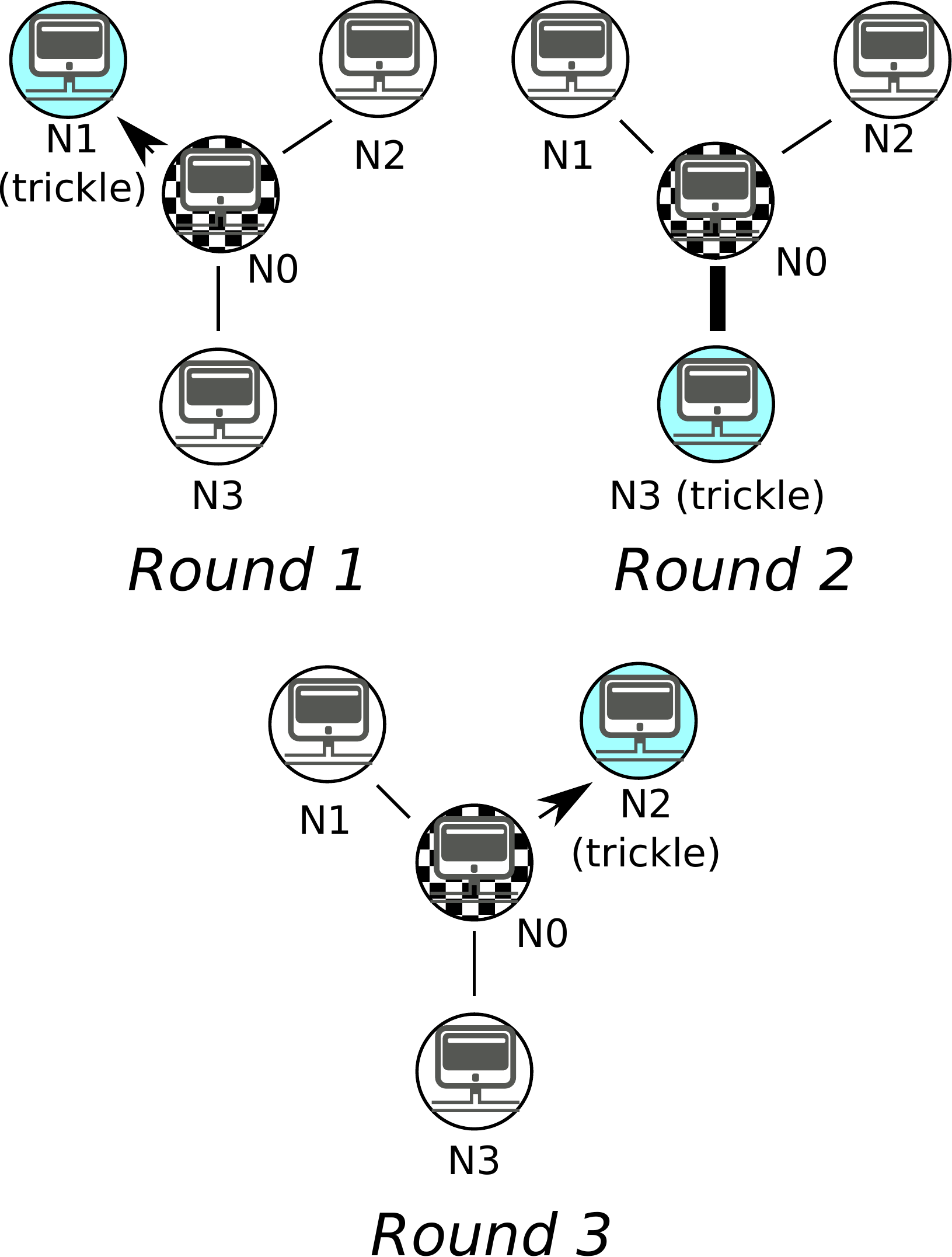}
\caption{Trickling of \code{ADDR} messages}
\label{fig:trickling}
\end{center}
\end{figure}

Consider an example on Fig.~\ref{fig:trickling}. Assume that node
\code{n0} gets an \code{ADDR} message with one address \code{$A_0$} from
node~\code{n3} and  that node \code{n0} schedules to forward it
to nodes \code{n1} and \code{n2} (i.e. these nodes are {\it responsible nodes}
for address \code{$A_0$}).
In round 1 node \code{n1} is chosen as a trickle node and the address is
forwarded to this node 
while the delivery to \code{n2} is still pending.
After 100 milliseconds in round 2 \code{n3} is chosen as the trickle
node thus no actual transmission happens at this stage.
After another 100 milliseconds in round 3 \code{n2} is chosen as the
trickle node and address \code{$A_0$} is finally sent to it.
Choosing a trickle node causes random delays at each hop during an
address propagation.

Finally for each connection, a Bitcoin peer remembers addresses that were
forwarded over this connection. Before a peer forwards an address, it
first checks if the same address was already sent over the connection.
This history is cleared every 24 hours.
An important note is that the history of sent addresses is kept per connection and not per
IP, i.e. if a Bitcoin peer reconnects, its history will be cleared.
The total number of addresses a Bitcoin peer can store is limited by 20480.
Whenever new addresses arrive at a peer they replace old ones (according to
specific rules which are outside of the scope of this paper). In addtition
when a peer receives a \code{GETADDR} messages it sends back 23\% of the number of 
addresses it stores but not more than 2500 addresses.

\paragraph{Peer discovery}
After the startup a Bitcoin peer discovers its own IP
addresses, which includes not only its network interfaces addresses
but also the IP address as it is seen from the Internet (in the majority
of cases for NAT users it resolves to an IP address of the peer's ISP).
In order to discover the latter, the peer issues a \code{GET}
request to two hard-coded web-sites which reply with the address.
For each address obtained by the discover procedure, the peer assigns
a score. Local interfaces initially get score 1, the external IP
address gets score of 4 (in case the external IP address coincides with
one of the local addresses the scores a summed).
When a client establishes an outgoing connection to a remote peer, they
first exchange \code{VERSION} messages and the client advertises its address
with the highest score. The remote peer then uses the addresses
propagation algorithm described above. The client repeats the same procedure
for the remaining 7 outgoing connections.

\paragraph{Transaction propagation}
Forwarding a transaction from one peer to another involves several
steps. First the sender transmits an \code{INVENTORY} message with
the hash of the transactions. Second, the receiver runs several checks
on the transaction and if the checks pass, it requests the actual
transaction by sending a \code{GETDATA} message. The sender then
transmits the transaction in a \code{TRANSACTION} message. When the receiver gets
the transaction he advertises it to its peers in an \code{INVENTORY}
message.

When a client generates a transaction he schedules\footnote{By
scheduling we mean that the node puts the transaction into the outgoing
queue  of the corresponding connection.} it for forwarding
to all of its  neighbours.
It then computes a hash of a value composed of the transaction hash and
a secret salt. If the computed hash has two last bits set to zero the
transaction is forwarded\footnote{More precisely the peer sends an
\code{INVENTORY} message with the hash of the transaction.} 
immediately to all the 8 entry nodes.
Otherwise a queue of a neighbour for outgoing transactions is
flushed when the neighbour becomes the trickle node (the same as with
\code{ADDR} messages).
Obviously $\frac{1}{4}$ of all transaction are forwarded immediately in
average.

When a transaction is received it
is scheduled for the delivery to all peer's neighbours as
described above.
As with \code{ADDR} messages, a Bitcoin peer maintains history
of forwarded  transactions for each connection. If a transaction was
already sent over a connection it will not be resent again.
A Bitcoin peer keeps all received transaction in a memory pool. If the peer
received a transaction with the same hash as one in the pool or in a block
in the main block chain, the received transaction is rejected.

\section{DISCONNNECTING FROM TOR}
\label{subsec:tor}
In this section we explain the first phase of our attack. We show how to prohibit the Bitcoin servers to accept connections via Tor and other anonymity services. This results in clients using their actual IP addresses when connecting to other peers and thus being exposed to the main phase of our attack, which correlates pseudonyms with IP addresses. This phase is quite noticeable, so a stealthy attacker may want to skip it and
deanonymize only non-Tor users.

In the further text we discuss Tor, but the same method applies to other anonymity services with minor modifications.
Briefly, the Tor network~\cite{dingledine2004tor} is a set of relays
(5397 for the time of writing) with the list of all Tor  relays publicly available on-line.
Whenever a user wants to establish a connection to a service through Tor,
he chooses a chain of three Tor relays. The final node in the chain is called
Tor Exit node and the service sees the connection as it was originated from this Tor Exit node.

To separate Tor from Bitcoin, we exploit the Bitcoin
built-in DoS protection. Whenever a peer receives a malformed message,  it increases the penalty score of the IP address from which the message
came (if a client uses Tor, then the message will obviously come from one of the Tor
exit nodes). When this score exceeds 100, the sender's IP is banned for 24 hours.
According to the \textsf{bitcoind} implementation, there are many ways to generate
a message which would cause penalty of 100 and an immediate ban, e.g.
one can send a block with empty transactions list (the size of such a message is
81  bytes).  
It means that if a client proxied its connection over a Tor relay and sent a malformed
message, the IP address of this relay will be banned.

This allows to separate any target server from the entire Tor network.
For that we connect to the target through as many Tor nodes as possible.
For the time of writing there were 1008 Tor exit nodes. Thus the attack requires
establishing 1008 connections and sending a few MBytes in data. This can be repeated
for all Bitcoin servers, thus prohibiting all Tor connections for 24 hours
at the cost of a million connections and less than 1 GByte of traffic.
In case an IP address of a specific Bitcoin node can be spoofed, it can be banned as well.

As a proof of concept we used the described method to isolate our Bitcoin node
from a set of Tor exit relays.

\paragraph{Possible countermeasures}
It is desirable to allow the Bitcoin peers to use Tor and still to keep some
blacklisting capability. We suggest making every connection
time- or computation-consuming to radically increase the attack
cost. For instance, any peer that initiates a connection might be
required to present some proof-of-work, e.g., a hash of its IP, the
timestamp, and the nonce that has a certain number of trailing zeros.
If we require 32 zero bits, then to separate a single peer from the Tor
network would cost about $2^{45}$ hash computations, which takes
several days on a modern PC. 

One may argue that some Bitcoin pools are powerful enough to afford that many hash calls. However, the vast majority of pool's computing power is contained in custom-built ASIC miners, which implement only a specific instance of SHA-256 and can not be reconfigured for another hash function, say, SHA-3. The exact fraction of GPU and CPU computing power is unknown, but at the time when these architectures were dominant, the total computing power was by several orders of magnitude smaller than now.


\section{LEARNING TOPOLOGY}
\label{sec:connectivity}
Suppose that we have ruled out the case that the Bitcoin users, which we deanonymize, use Tor. Now we target clients, i.e. nodes that do not accept 
incoming connections, but have 8 outgoing connections to \emph{entry nodes}.
In this section we show how to learn these entry nodes. 

\label{subsec:revealconnectionsclients}
The method is based on the fact that
whenever a client $C$ establishes a connection to one of its entry nodes,
it advertises its address $C_a$ as it is seen from the Internet (see
section~\ref{sec:background}). If the attacker is already connected to an entry node, with some probability (which
depends on the number of the attacker's connections)  the
address $C_a$ will be forwarded to him. This suggests the following strategy:
\begin{enumerate}
\item Connect to $W$ Bitcoin servers, where $W$ is close to the total number of servers.
\item For each advertised $C_a$, log the set $E'$ of servers that forwarded $C_a$ to attacker's machines and designate it as the entry node subset $E_{C_a}'$.
\end{enumerate}
There are two problems with this method. First, the entry node might send the client's address to some non-attacker's peer.
Second a client does not connect to all his entry nodes simultaneously, but there is a time gap between connections. In both cases, the advertised address reaches
attacker's machines via peers that are not entry nodes, which yields false (noisy) entries in $E_{C_a}'$. 

\paragraph{Noise reduction technique}
Our strategy of filtering noise assumes that either the client's IP was already used in the Bitcoin network, which is quite common for the clients behind NAT
or the client's public IP is contained in a known list of IP addresses (e.g. within an IP range of a major ISP) which an attacker can use.
If an attacker knows  $C_a$, he restricts its propagation using the  following fact:
\begin{itemize}
\item If the address had already been sent from $A$ to $B$, it will not be forwarded over
this connection again;
\end{itemize}
This suggests broadcasting $C_a$ (or all the addresses under investigation) to all servers we are connected to.
We suggest repeating this procedure every 10 minutes (see details below), though there could be other options.
The adversary expects that when the client reconnects, the entry nodes will forward $C_a$ to him, and even if they don't, the address propagation will stop before it reaches the adversary via a non-entry node.

Eventually the attacker obtains the fraction $p_{addr}$ of client's entry nodes. The exact value of $p_{addr}$ depends on the number of attacker's connections,
and it is computed for some parameters in Section~\ref{sec:an-num}. For instance, if an attacker establishes 35 connections to each potential entry node, which all had 90 connections beforehand, then he 
identifies
4 entry nodes out of 8 on average.

Here are some details. 
When the attacker advertises the
$C_a$, each  Bitcoin
server chooses two {\it responsible nodes} to forward the address.
The attacker than establishes a number of connections to each server
in the network hoping that her nodes will replace some of the
responsible nodes for address $C_a$.
When client \code{$C$} connects to one of its entry nodes \code{e1},
it advertises its address. If one of attacker's nodes replaced
one of the responsible nodes, then the attacker will learn that
client $C$ might be connected to node \code{e1}. If the responsible
nodes did not change address $C_a$ will not be propagated
further in the network.

Since the attacker  advertised $C_a$
to node \code{e1}, responsible nodes of \code{e1} might be replaced by some
non-attacker nodes and the attack might fail. In Section~\ref{sec:analysis} we show that the probability of this event is actually quite low
given that the attacker re-sends its list of addresses frequently enough.

\section{DEANONYMIZATION}
\label{sec:deanon}

We have prohibited Bitcoin servers from accepting Tor connections 
and showed how to find the entry nodes of clients. Now we describe the main phase of the deanonomyzation attack.

The main phase consists of four steps:
\begin{enumerate}
\item Getting the list $S$ of servers. This list is regularly refreshed.
\item Composing a list $\mathbf{C}$ of Bitcoin clients for deanonymization.
\item Learning entry nodes of clients from $\mathbf{C}$ when they connect to the network.
\item Listening to servers from $S$ and mapping transactions to entry nodes and then to clients.
\end{enumerate}

Eventually we create a list $\mathbf{I} = \{(IP,Id,PK)\}$,
where $IP$ is the IP address of a peer or its ISP, $Id$ distinguishes clients sharing the same IP, and $PK$ is the pseudonym used in a transaction (hash of a public key). Let us
explain the steps in detail.

\paragraph{Step 1. Getting the list of servers} 
This phase of the attack is rather straightforward. An attacker first collects the entire list  of peers by quering all known peers with a 
\code{GETADDR} message. Each address $P$ in the response \code{ADDR} message
can be checked if it is online by establishing a TCP connection and
sending a \code{VERSION} message. If it is, $P$ is designated as a server.
An attacker can initiate the procedure
by querying a small set of seed nodes and continue by querying the newly received
IP addresses. 
The adversary establishes $m$ connections to each server (we suggest 50 for the size of the current Bitcoin network).

\paragraph{Step 2. Composing the deanonymization list}
The attacker selects a set $\mathbf{C}$ of nodes whose identities he wants to reveal. The addresses may come from various sources. The attacker might take IP ranges of major Internet
service providers, or collect addresses already
advertised in the Bitcoin network. Finally, she might take some entries from the list of peers she obtained at Step 1.

\paragraph{Step 3. Mapping clients to their entry nodes}
Now the attacker identifies the entry nodes of the clients that are connecting to the network. Equipped with the list $C$ of addresses, the attacker runs the procedure described
in Section~\ref{sec:connectivity}. Let us estimate how many entry nodes are needed to uniquely identify the client.

Let us denote  the set  of entry nodes for $P$ by
$E_P$. We stress that it is likely that $E_{P_1} \neq E_{P_2}$ even if $P_1$ and $P_2$ share the same IP address.   For each $P$ advertising its address in the network the attacker obtains a
set of  $E_P' \subseteq E_P$. Since there are about $8\cdot 10^3$ possible entry nodes out of $10^5$ total peers (servers and clients together), the collisions in $E_P'$ are unlikely if every tuple has at least 3 entry nodes:
$$
\frac{10^5\cdot 10^5}{(8\cdot 10^{3})^3} \ll 1.
$$
Therefore, 3 entry nodes uniquely identify a user, though two nodes also do this for a large percent of users.

An attacker adds $E_P$ to its database and proceeds to Step 4.

\paragraph{Step 4. Mapping transactions to entry nodes}
This step runs in parallel to steps 1-3. Now an attacker tries to correlate the transactions appearing in the network with sets $E_P'$ obtained in step 2.
The attacker listens for \code{INVENTORY} messages with transaction hashes received over all the connections
that she established and for each transaction $T$ she collects $R_T$ ---  the first $q$ addresses of
Bitcoin servers that forwarded the \code{INVENTORY} message. She then compares  $E_P'$ with $R_T$ (see details below), 
and the matching entries suggest pairs $(P,T)$. In our experiments we take $q=10$. 

There could be many variants for the matching procedure, and we suggest the following version.
\begin{itemize}
\item The attacker composes all possible 3-tuples from all sets $E_P'$ and looks for their appearances in $R_T$. If there is a match, he gets a pair $(R,T)$;
\item If there is no match, the attacker consider 2-tuples and then 1-tuples. Several pairs $\{(P_i,T)\}$ can be suggested at this stage, but we can filter them with later transactions.
\end{itemize}

We made a bunch of experiments and collected some statistics to estimate the success of the attack. Even the first step is quite powerful.
In our experiments on the testnet we established 50 connections to each server, obtained 6 out
of 8 entry nodes on average, and the 3-tuples were detected and linked to the client
in 60\% of transactions (Section~\ref{sec:experiments}). In the real network, where we can establish fewer connections on average, our pessimistic
estimate is 11\% (Section~\ref{sec:analysis}), i.e. we identify 11\% of transactions.

The 2-tuples may suggest several pairs. Each client has $2^5$ 2-tuples of entry nodes, whereas the top-10 suggests $2^{5.5}$ 2-tuples. The matching probability is $2^{-26}$, which implies that 
the top-10 suggests $2^{16.5+10.5-26} = 2$ clients on average on the 2-tuple rule. The probability for the right client to be detected we estimate as $0.28$ in Section~\ref{sec:analysis}, which means that each transaction suggests two clients, but only in 28\% cases the right one is among those two.

\begin{figure}[h]
\begin{center}
\includegraphics[scale=0.8]{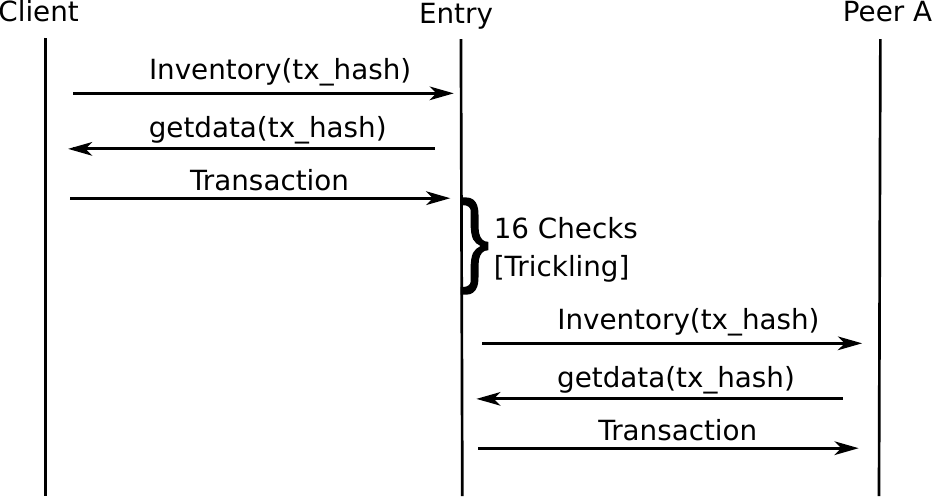}
\caption{Steps necessary to forward a transaction}
\label{fig:tx-forwarding}
\end{center}
\end{figure}

\paragraph{Remark 1}
Step 4 of the attack depends on that some entry nodes of a client are among the first to
forward the \code{INVENTORY} message with the transaction's hash. 
The intuition behind it
is that it takes a number of steps for a transaction to propagate to the next hop.
Fig.~\ref{fig:tx-forwarding} shows steps that are required for a transaction
to be propagated over two hops and received at peer \code{A}. When a transaction is
received by a node it first runs a number of checks and then schedules the
transmission. The actual transmission will happen either immediately (for 25\% of transations)
or with a random delay due to trickling (see Section~\ref{sec:background}).
The time needed for an \code{INVENTORY} message  to be forwarded to the attacker's node through
node \code{Entry} is the sum of propagation delays of 4 messages (2x\code{INVENTORY},
1x\code{GETDATA}, 1x\code{TRANSACTION}) plus the time node \code{Entry} needs to run 16 checks
and possibly a random trickling delay. On the other hand the time needed for the same 
\code{INVENTORY} message to be forwarded to the attacker's node through peer \code{A}
consists of 7 messages (3x\code{INVENTORY},
2x\code{GETDATA}, 2x\code{TRANSACTION}), 32 checks, and two random delays due to trickling.
Finally since the majority of connections to a peer are coming from
clients, one more hop should be passed before the transaction reaches an attacker's node
through a wrong server.
Measurements of transaction propagation delays are given in Appendix C.

Based on this we expect that if a transaction generated by a client is forwarded to
the entry nodes immediately, the entry nodes will be the first nodes to forward the
transaction.
In case when the transcation was sent sequentially with 100 ms between transmissions
we still expect a fraction of entry nodes to be among the first 10 to forward corresponding
\code{INVENTORY} message to one of the attacker's nodes.
This fraction obviously depends on the propagation delay between Bitcoin peers.
The higher the propagation delay the less significant becomes delay of 100 ms in
trickling. For example if the propagation delay is 300 ms between the client and each entry node
it's likely that 3 entry nodes
will be among the first to forward the \code{INVENTORY} message (given 
that the attacker has enough connections to Bitcoin servers).

\paragraph{Remark 2}
The attack presented in this section requires from an attacker only to
be able to keep a significant number of connections to Bitcoin servers
without sending large amount data. In order to make the attack less 
detectable an attacker might decide to establish connection to a given
Bitcoin server from different IP addresses, so that all connection look
like they came from different unrelated clients. The same set of IP addresses can
be used for different servers.

\paragraph{Remark 3}
The technique considered in the section provides unique identification of
Bitcoin clients for the duration of a session, and thus if a client makes
multiple transactions
during one session they can be linked together with very high probability. Note
that this is done even if the client uses totally unrelated public keys/Bitcoin
wallets, which have no relation in the Bitcoin transaction graph and thus such
linkage would be totally unachievable via transaction graph analysis~\cite{Meiklejohn,ron2013quantitative}.
Moreover we can easily distinguish all the different clients even if they come
from the same ISPs, hidden behind the same NAT or firewall address.

\paragraph{Countermeasures}
As a possible countermeasure against client de-anonymization we propose to
change the client octet after every transaction and add some random delay after the
transaction (to avoid timing linkability attack). This will remove likability
of transactions and will also prohibit distinguishing of different clients from
the same ISP. This however will not prevent the attacker from learning the ISP
of the client.

\section{EXPERIMENTAL RESULTS}
\label{sec:experiments}

As a proof of concept we implemented and tested our attack on the Bitcoin
testnet. We did not perform a deanonymisation attack on real clients for
ethical reasons.
For our experiments we built our own Bitcion client,
which included functionality
specific for our attack -- sending specific Bitcoin messages on request or
establishing various numbers of parallel connections to the same Bitcoin server, etc.
When imitating clients we used the main Bitcoin client.
In order to periodically get the list of all running Bitcoin servers we used 
an open source crawler~\cite{bccrawler}.

For the time of experiments (May 2014) the number of running Bitcoin
servers in the testnet fluctuated between 230 and 250, while the estimated average degree 
of the nodes was approximately 30.
In our experiments we were imitating several different users connecting to
the testnet from the same ISP's IP address and from different ISP's at different times.
As an attacker we added 50 additional connections to each Bitcoin server.
For each experiment in the first phase of the attack we propagated clients' addresses in the testnet
10 minutes before they started to send their transactions. 
In total we (as clients) sent 424 transactions.

\begin{figure}[h]
\begin{center}
\includegraphics[scale=0.6]{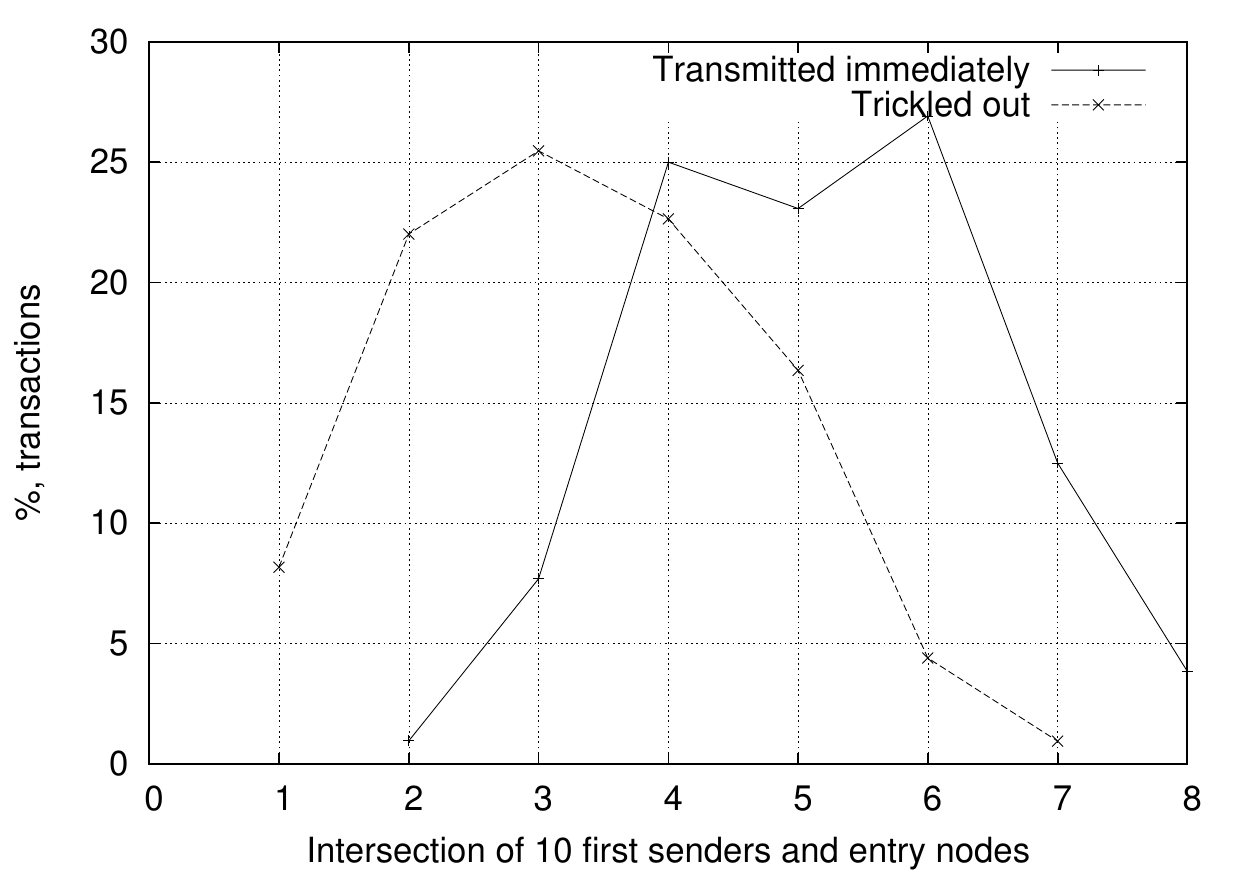}
\caption{Intersection of top-10 senders and entry nodes}
\label{fig:top10-intersect-with-entries-combined}
\end{center}
\end{figure}

In the first experiment we confirm our expectations that transactions are first
forwarded by entry nodes and analyse the number of entry nodes that were among the first
10 to forward the transactions (i.e. we assume that the attacker correctly identified
all entry nodes). We splitted all transactions into two sets:
the first set contains 104 transactions, which were forwarded to the entry nodes
immediately; the second set contains all other 320 transactions (i.e. for which trickling
was used).
Fig.~\ref{fig:top10-intersect-with-entries-combined}
shows the number of entry nodes that were among the first 10 to forward the 
transaction to the attacker's nodes for these two sets.
As expected if a transaction was immediately forwarded to all entry nodes
the attacker was able to ``catch'' three or more of them in 99\% of cases.
In case of transactions from the second set, the attacker was able to ''catch''
3 or more entry nodes in 70\% of cases.
We also observed that for the majority of transactions the first two nodes to forward
the transaction to the attacker were the entry nodes.

In the second experiment we executed all steps of the attack.
In our experiment each client was successfully uniquely identified by
his own set of entry nodes and
on average we identified 6 entry nodes for each client.
Assuming that 3 entry nodes is enough for unique identification of a client
we correctly linked 59.9\% of all transactions to the corresponding IP
address by matching entry nodes of clients and first 10 Bitcoin servers
which forwarded the transaction.
We correctly glued together all transactions of the same client which were
made during one session.

In a bit more conservative setting we added only 20 additional nodes in which case
we successfully deanonymised 41\% of our transactions.

\section{ANALYSIS}
\label{sec:analysis}
The success rate of the attack presented above depends on a number of
parameters, among which the most important is the fraction of attacker's
connections among all the connections of client's entry nodes.
The fewer the number of connections of entry nodes are, the more
connections the attacker can establish and the higher chance is to deanonymise the client.
In this section we analyze each step of the attack and
compute success rates for some parameter sets.

\subsection{Number of connections to servers}\label{sec:an-num}
Both mapping client to entry nodes and mapping entry nodes to transactions
depends on the number of connections the attacker can establish to the
Bitcoin servers. 
Assuming the entry node had $n$ connections and the attacker added
$m$ new connections, thus the total number of connections is $N=n+m$,
the probability to receive the address at the
first hop is $p_{addr}(n,N) = 1-\frac{n}{N}\cdot \frac{n-1}{N-1}$.
For a transaction which was not
forwarded immediately to the peer's neighbours the probability that one
of attacker's nodes is chosen as trickle node in the first round is $p_{tx} = \frac{m}{N}$.
For $n=50$,  $m=50$,
$p_{addr}=0.75$ and $p_{tx}=0.50$. For $n=90$, $m=35$, $p_{addr}=0.49$
and $p_{tx} = 0.28$. The number of connections that the adversary can establish
is limited by the total number of 125 connections a Bitcoin peer can have
by default.

\begin{figure}[h]
\begin{center}
\includegraphics[scale=0.6]{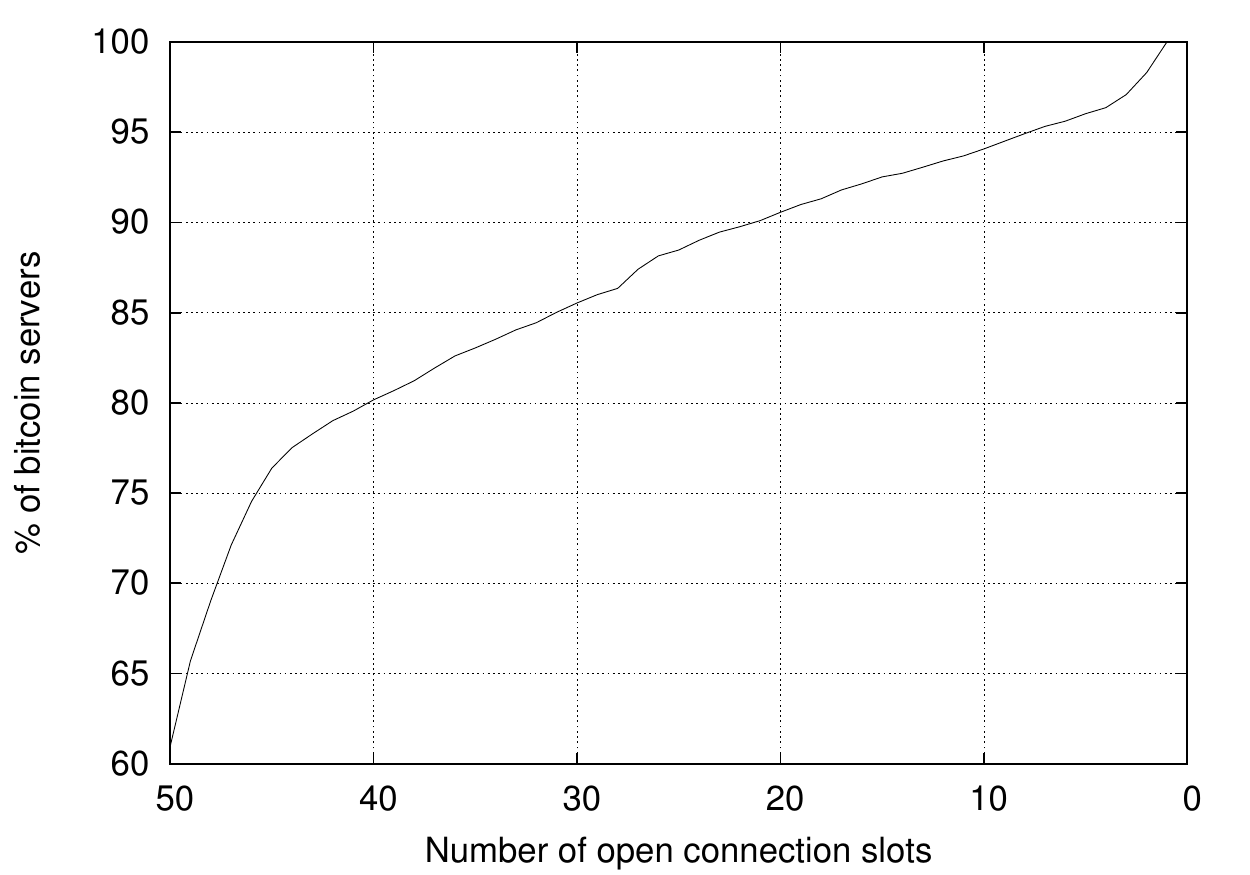}
\caption{Distribution of open slots}
\label{fig:free-slots}
\end{center}
\end{figure}

In order to see how many open connection slots Bitcoin peers have we
conducted the following experiment in April 2014.
For each Bitcoin server that we found we tried to establish 50
parallel connections and
check the actual number of established connections\footnote{We did not try establish more
than 50 connections in order not to degrade the Bitcoin network performance.}.
Fig.~\ref{fig:free-slots} shows
the distribution of number of established connections.
The experiment shows that 60\% of peers allow 50 connections or more, and
80\% of Bitcoin peers allowed up to 40 connections.
Note that even if sufficient number of connection cannot be
established to a Bitcoin peer immediately
they can be established in longer term since many Bitcoin clients will
eventually disconnect and thus allow new connections (according to an example
disconnection rate as shown in Fig.~\ref{fig:runaways} it might take several hours,
but once an attacker got the required number of connections she can keep them as long as 
needed).
Also note that Bitcoin servers allow any number of
connections from a single IP address.

Finally the attacker does not send much traffic over the established
connections but rather listens for messages. Incoming traffic is normally
free of charge if one rents a server.
Thus in spite of the large number of connections that an attacker
needs to establish the attack remains very cheap.

\subsection{Estimating false positives}
Assume that some of the steps of that attack fail.
Then the first 10 peers to report the transaction to the attacker will be
some random Bitcoin peers.
If there is no 3-subset of these 10 that match some entry node
set, then such a transaction is marked as unrecognized by an attacker.
The probability that nodes accidentally match any
set of Bitcoin entry nodes (we call this a collision) for a given
transaction is
$$
p_c = \binom{10}{3}\times \binom{10}{3}\times \left(\frac{1}{N}\right)^3
$$

\noindent where $N$ is the number of Bitcoin
servers in the network. Given that there are about
8000 Bitcoin servers and 100,000 Bitcoin clients,  the number of incorrectly assigned
transactions is negligible.

We now estimate the probability that an attacker adds a wrong entry node
to the set of entry nodes of a particular client
(we recall that according to the address propagation mechanism after receiving 
an address a peer forwards it to only two randomly chosen {\it responsible nodes}).
For this to happen, one or more entry nodes should forward the client's
address $C_a$ over one of non-attacker's connections, whence (since the attacker
periodically propagates the client's address)
at least one of responsible nodes for address $C_a$ should change on
an entry node after the attacker last propagated $C_a$.

\begin{figure}[h]
\begin{center}
\includegraphics[scale=0.6]{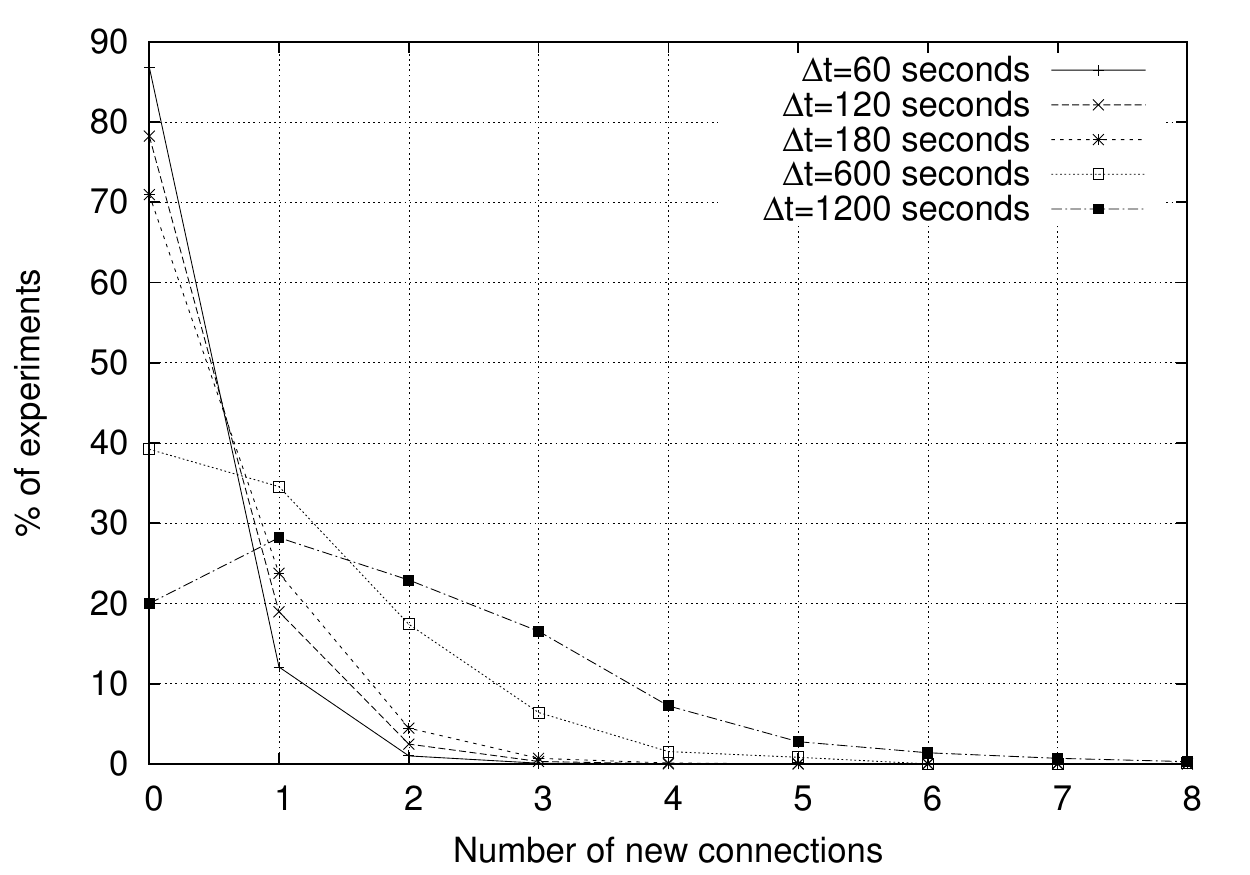}
\caption{Probability density of number new connections}
\label{fig:newcomers}
\end{center}
\end{figure}

In order to estimate this probability we collected statistics
from our Bitcoin peer for 60 days from March 10 till May 10 2014.
We collected information about 61,395 connections in total.
Assume that the attacker propagated $C_a$ at time $t_0$,
the probability that a responsible node will be different at time $t_1=t_0+\Delta t$
depends on the number of new connections the entry node has at $t_1$ and
number of nodes that disconnected since $t_0$.
Fig.~\ref{fig:newcomers} shows probability density function of the
number of new connections (i.e. the incoming connections rate) for different values of $\Delta t$.

\begin{figure}[h]
\begin{center}
\includegraphics[scale=0.6]{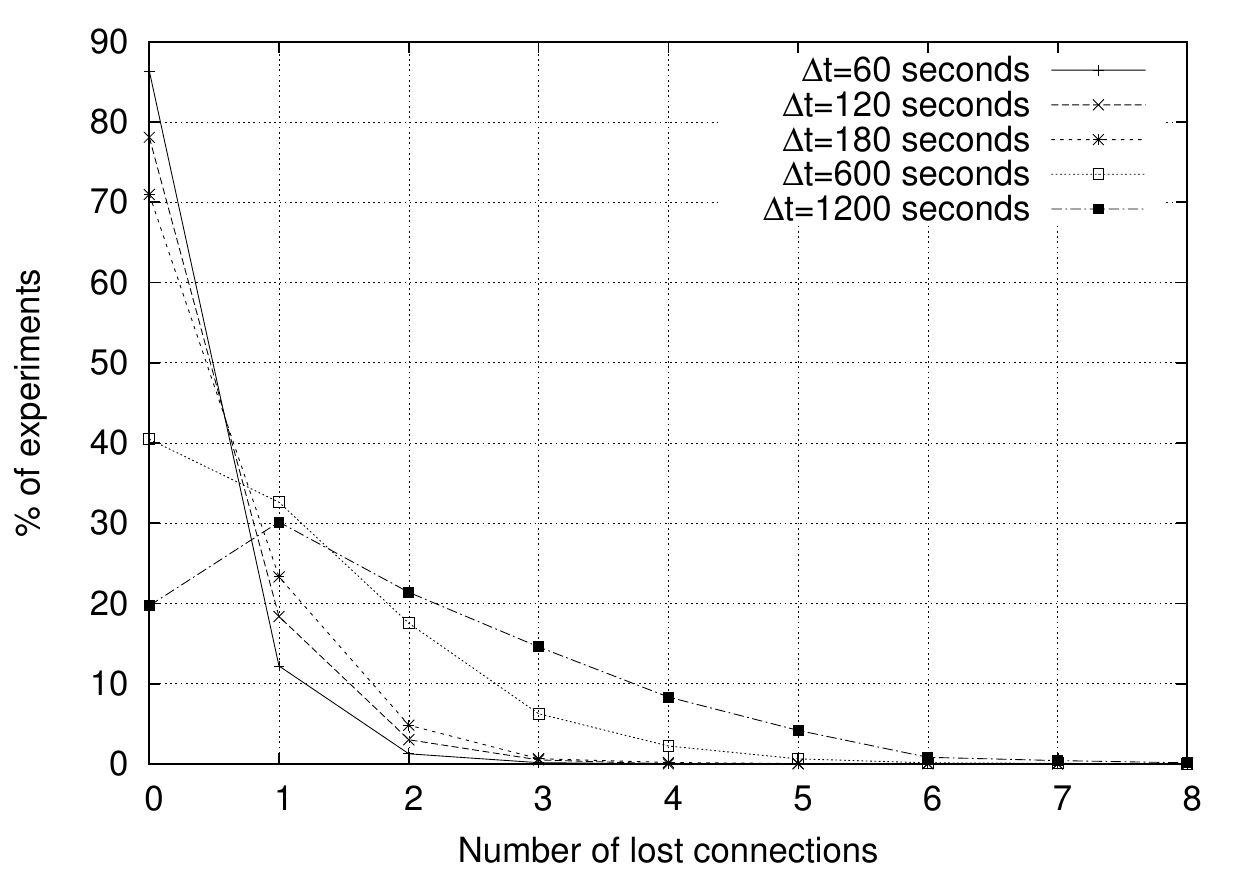}
\caption{Probability density of number lost connections}
\label{fig:runaways}
\end{center}
\end{figure}

Fig.~\ref{fig:runaways} shows probability density function of the
number of disconnection (i.e. connection close rate) for different values of $\Delta t$.

\begin{figure}[h]
\begin{center}
\includegraphics[scale=0.6]{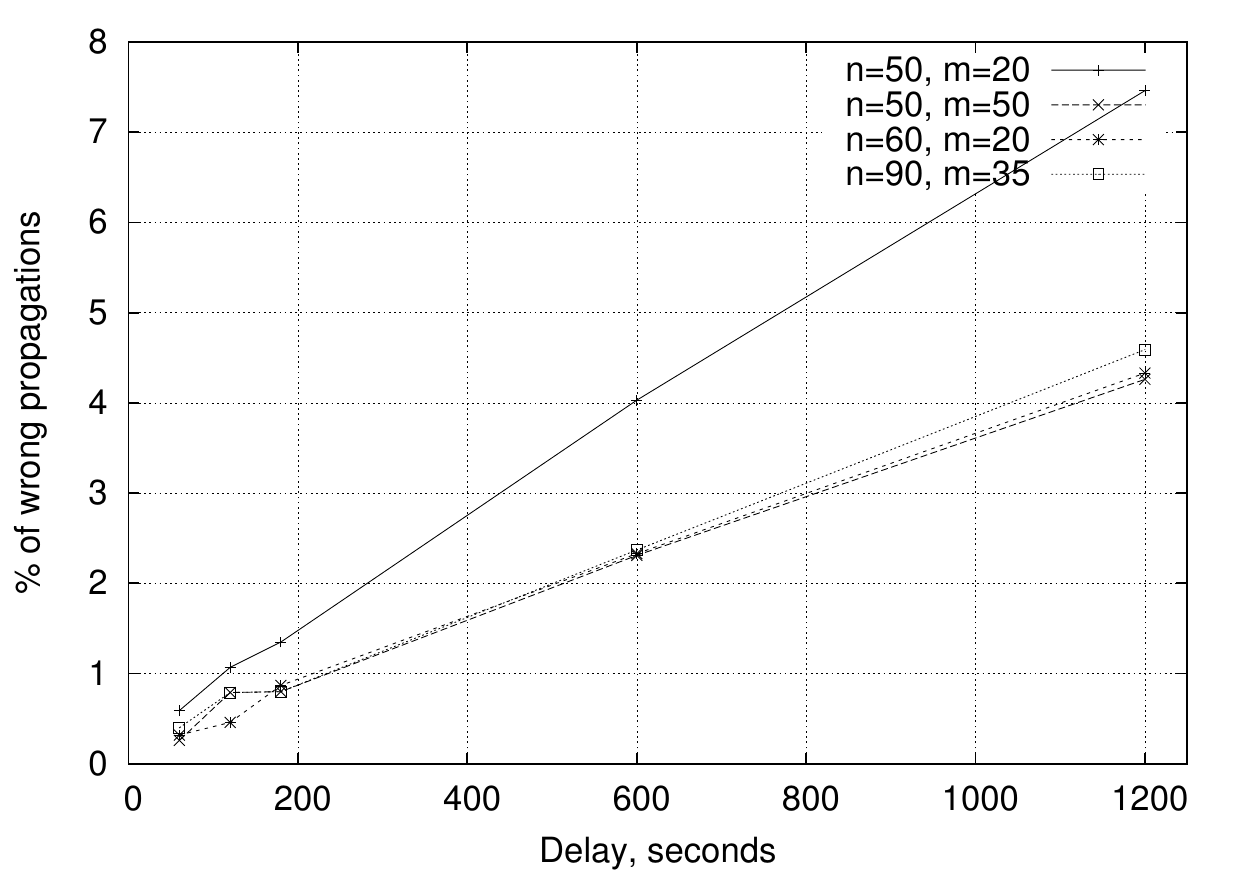}
\caption{Percentage addresses forwarded by entry node over non-attacker
connections}
\label{fig:entry-fail}
\end{center}
\end{figure}

We use these distributions to simulate the address
propagation and calculate the probability that the client's address is
forwarded by an entry node over a non-attacker's link after time
$\Delta t$ after the attacker sent this address over the network.
We obtained probabilities for different number of attacker's and
non-attackers's connections and for each connection setting and each
$\Delta t$ we executed 10,000 runs of the model.
Fig.~\ref{fig:entry-fail} shows the obtained probabilities. The number of
attacker's connections is denoted by $m$ and the number of
non-attacker's connections by $n$.

As expected, the more connection a node has the less probable that the
responsible nodes for an address will change after $\Delta t$. Another
observation is that the probability of a node to forward the client's
address over one of the non-attacker's connections depends on the total
number of connections rather than on the fraction of attacker's connections.
From Fig.~\ref{fig:entry-fail} we conclude that resending client addresses
over the Bitcoin network every 10 minutes seems to be a reasonable choice.
Also note that even if a client's address was forwarded over a
non-attacker's link, the further propagation of the address will likely
stop at the next hop.

\subsection{Overall success rate}\label{sec:rate}
The success rate $P_c$ of the attack depends on a number of characteristics of the real network.
We propose the following method to estimate it. First, we assume that the attacker establishes all possible connections to Bitcoin servers. From the data used in Figure~\ref{fig:free-slots}, we
estimate the average value $p_{addr}^{Avg}$ of the parameter $p_{addr}$. We did not establish more than 50 connections to avoid overloading servers, and we take a pessimistic estimation that
50 is the maximal number of attacker's connections. This yields
$$
p_{addr}^{Avg} \approx 0.34.
$$
Then we assume that both the testnet and the mainnet exhibit similar local topology so the probabilities $\mathbb{P}_3(L)$ for the number $L$ of entry nodes being in top-10 are almost the same (Figure~\ref{fig:top10-intersect-with-entries-combined}).  We calculate the probabilities $\mathbb{P}_1(R)$ for the number $R$ of entry nodes being detected out of 8 as a function of
$p_{addr}^{Avg}$. Then we compute the total probability 
that the adversary detects at least $M3=$ nodes among those appeared in top-10, and we get the following estimation (details in Section~\ref{sec:succ-det}):
$$
\mathbb{P}_{success}(3)\approx 0.11.
$$
When we restrict to 2-tuples, the success rate increases to $0.35$.

In the testnet we managed to achieve $p_{addr}^{Avg} = 0.86$ and the success rate for $M=3$ being close to 60\%. 
An attacker may achieve such high rates if he first saturates servers' connections and then gradually replaces
the expired connections from other nodes with his own ones. However, this may cause 
degradation of QoS as some clients will be unable to connect to all their entry nodes.

Thus a careful attacker that followes the 3-tuple rule only and establishes 50 connections at maximum to each server can  catch about 11\% of transactions generated by clients. Given 70,000
transactions per day, this results in 7,700 transactions per day.
This also means that a user needs to send 9 transactions in average in
order to reveal his public IP address.

\section{ALTERNATIVE REALITY}
\label{sec:alt-real}
In this section we show how to create and maintain an alternative block chain while keeping the Bitcoin protocol, existing wallets and transactions untouched. This procedure can be used by the Bitcoin community  
if the current difficulty becomes too high and impossible to sustain. It can also be an attack vector for a malicious admininstrator of the Bitcoin code project.

For motivation, suppose that the mining stops being profitable so that the most powerful miners quit the network in order to stop paying large electricity costs. As a result, the total computational power drops
and the network outputs blocks at a slower rate. 

The Bitcoin protocol is quite reactive to the rise of the hashrate, but has large latency in the opposite case. The maximum difficulty drop is the factor of 4 and requires the 2016 blocks to be produced within at least 8 weeks. Not only it is a long recovery period, but also the network might be so slow that the necessary number of blocks might not be produced at all. Our solution creates an alternative reality with lower difficulty and the same transaction history within a small timeframe.

\paragraph{Block construction rules}
It may happen that distinct miners create blocks almost simultaneously,
which is called a \emph{fork}. In this case the network temporarily splits
into two parts, which try to generate a valid block at their own end of
the fork. When a new block is found by either part, a \emph{higher
difficulty principle} applies: the chain with higher total
difficulty\footnote{The real formula is complicated, but the total
computational complexity of generating the chain is a good
approximation.} is accepted, and the one with lower is discarded. As a
result, a new block at either end of the fork yields a chain with
higher difficulty, and the chain is accepted by all the peers. Due to
this short term uncertainty it is recommended to wait at least 6 blocks
(i.e. about 60 minutes) as a confirmation that the transaction became
part of the block chain. To further fix the block chain, the
administrators of \textsf{bitcoind} routinely hard-code hashes of some blocks
into the client (currently 12 \emph{checkpoint} blocks, on average
every 25,000-th, are hard-coded) code.

The Bitcoin protocol aims to sustain a constant rate of block
production. Every 2016 blocks (about two weeks) the difficulty is
recalculated so that blocks are generated every 10 minutes. The exact
rules are as follows:
\begin{itemize}
  \item For block $X_i$, $i \neq 2016k$, the difficulty is equal to
        that of $X_{i-1}$;
  \item For block $X_i$, $i=2016k$, we extract the time $T_2$ of block
        $X_{i-1}$ and the time $T_1$ of block $X_{i-2016}$. Let the
  time difference $\Delta T = T_2 - T_1$ be expressed in days.
  Then the difficulty is multiplied by $\frac{14}{\Delta T}$.
  The protocol also enforces that the multiplier can not exceed
  4 or be smaller than $0.25$.
\end{itemize}

Bitcoin enforces a number of additional restrictions to discourage
malicious miners to play with timestamps for their own benefit.
The following two rules are important for us:
\begin{itemize}
  \item The timestamp of $X_i$ can not be older than the median
        (i.e., the middle element of the sorted array) of 11 previous
  timestamps.
  \item The difficulty $d_i$ of $X_i$ can not be lower than the
        hypothetical difficulty yielded by reducing the last
  checkpoint difficulty by the factor of 4 every 8 weeks,
  i.e. the minimal difficulty that is possible if the network
  slows down.
\end{itemize}

\paragraph{Alternative block chain}

Alternative chain is constructed as follows. First, we select the first block $X_i$ after the last checkpoint such that  2016 divides $i$: $i=2016k$, i.e. the difficulty is recomputed at this point. We create an alternative block with the same transactions but the date changed to the current date, which will decrease the difficulty of the subsequent blocks by the factor of 4. The next 2015 blocks  we create with arbitrary times, possibly immediately one after another, with $X_{i+1}$  and later possibly close to $X_{i-1}$.

The date of  block $X_{i+2016}$ we set again to the current time so that the total difficulty would drop as much as possible. The next blocks will be again older than $X_{i+2016}$. We
 repeat this procedure further and further until the resulting difficulty contradicts the difficulty of the checkpoint. If $T_c$ is the  date (in days) and $Q_c$ is
the difficulty of the last
checkpoint the client has in memory,  $T$ is the date  and $Q$ is the difficulty of the processed block, than the lower bound is
$$
Q \geq \frac{Q_c}{2^{\frac{T-T_c}{28}}}.
$$

Currently, a new checkpoint is added every 25000 blocks, which amounts to the period of about 140 days with the current difficulty increase rate. Therefore, the difficulty may drop by the factor of $2^{10}$ compared to the previous checkpoint.

As a result, we create an alternative reality where all the participants have the same balance. However, the new chain is not accepted by clients since it would have the smaller total difficulty compared to the original chain. To finish the switch to the new reality, a new checkpoint must be chosen on the new chain and distributed among the clients. Alternatively, high-difficulty blocks can be added to the beginning of the alternate chain to make it more difficult than the original one. Higher granularity achieved by lower difficulty at the end of the alternative chain would allow to surpass the original chain even if the last checkpoint is not set.

Let us estimate the amount of computational power needed for this operation. Suppose that we have waited for 25000 blocks after the last checkpoint. This occurred in Dec 15th, 2013 with the block 275000, with the checkpoint block 250000 generated on August 3d, i.e., 134 days before. It has difficulty smaller by the factor of 30, let us denote it by $D$. In turn, the difficulty in our new history can
be even lower by approximately $2^{\frac{134}{28}} \approx 30$. To obtain that, we would have to create 2016 blocks with difficulty  $D/4$ and 2016 blocks with difficulty $D/16$. The other
23000 blocks must be created
with difficulty $D/30$. This amounts to about 1400 blocks with difficulty $D$, or less than 50 blocks with current difficulty. This means that a mining pool with only 10\% of the network
computational power would need only 3 days to make this happen.

\begin{figure}
\includegraphics[scale=0.3]{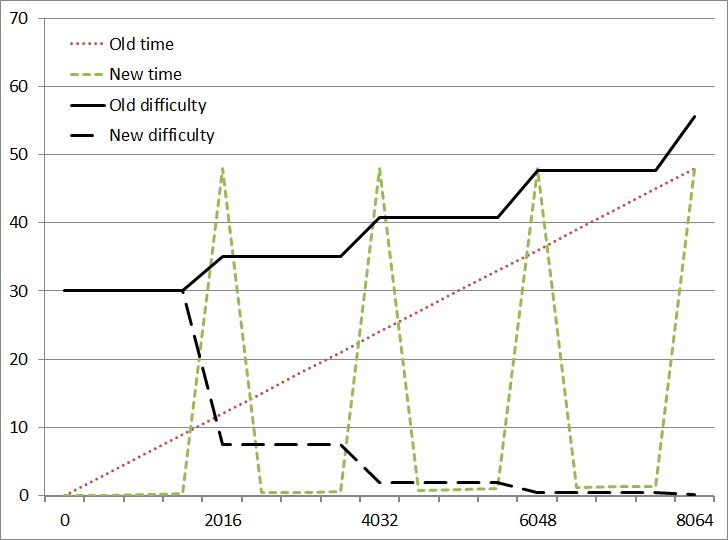}
\caption{Schematic view of the alternate reality creation over 8064 blocks. Vertical axis denotes the block index, horizontal axis denotes the timestamp (in days) and the difficulty.}
\end{figure}

\section{FURTHER LEARNING THE TOPOLOGY}
In this section we continue learning the topology of the Bitcoin network and show how to learn
server-to-server connections.
Bitcoin peers share information only about other peers they
know, but not about their direct connections.
In this section we first provide a method to estimate a node's degree
(the number of connections) and then show how to determine to which
servers it is connected to.

Some of the attacks exploit the following feature of address propagation mechanism.
Each forwarded
address is accompanied with a timestamp. If this timestamp is older than
10 minutes, the address is not retransmitted anymore. Thus
in order to avoid false positives in some attacks described below an attacker
should set the timestamps of the marker addresses to values close to 10 minutes to the past.

\subsection{Estimating number of connections}
\label{subsec:numofconnections}
Our first method is based on the way a Bitcoin peer forwards addresses
received in \code{ADDR} messages (see section \ref{sec:background} for
more details).
Suppose that a Bitcoin node \code{A} is connected to $k$ nodes
$\code{P_1,P_2,\dots,P_k}$. We show now how to estimate $k$.
First, we connect our node $X$ to peer
\code{A} and send it a set of fake\footnote{By fake me mean that no
Bitcoin nodes are running at these addresses.} marker addresses
$S=\{ip_1,ip_2,\dots,ip_n\}$ in portions of 10 addresses per \code{ADDR}
message. At the same time we listen (either on the same connection or on a
separate connection) for received marker addresses. As instructed by the
Bitcoin protocol, node \code{A} forwards marker addresses to its peers
including $X$. As the number of marker addresses increases the number of
addresses received by $X$ converges to $\frac{2}{1+k}$ if marker addresses
are considred reachable by \code{A} or $\frac{1}{1+k}$ otherwise, from which
we estimate $k$. We increase the accuracy by 1) running several listening
nodes, 2) repeating the same experiment several times. We can reuse the
same marker addresses: when we reconnect our listening nodes, peer \code{A}
revokes their histories and allocates new data structures for this nodes.
We note that all connections can be established from the same IP address.

We implemented this method and carried out several experiments.
Our target nodes have 10, 30, 70, or  100 connections.
For different number of connections we used different number of marker addresses
and different number of listening connections (2, 3, 7, and 10 correspondingly).
For each number of connections we conducted a series of experiments; Table
~\ref{tab:numofconnections1} shows five random runs for each series.

\begin{center}
\begin{table*}[ht]
  {\small
    \hfill{}
    \begin{tabular}{ | c *{6}{| c } | c | }
    \hline
    \multirow{2}{*}{Real connections ($k$)} & \multirow{2}{*}{Markers sent} & \multicolumn{6}{c|}{Estimates using our formula} \\
    \cline{3-8}
     & & Try \#1 & Try \#2 & Try \#3 & Try \#4 & Try \#5 & Average \\
    \hline
    10 & 500  & 10.69 & 9.57 & 9.34 & 10.6 & 11.38 & 10.32 \\
    \hline
    30 & 1000  & 31.92 & 30.88 & 35.47 & 36.26 & 30.48 & 33 \\
    \hline
    70 & 1000  & 72.92 & 76.84 & 70.65 & 64.16 & 77.36 & 72.39 \\
    \hline
    100 & 2000  & 102.63 & 109.12 & 104.27 & 103.28 & 95.66 & 103 \\
    \hline
    \end{tabular}
    \hfill{}
  }
  \caption{Estimation of number of connections}
  \label{tab:numofconnections1}
\end{table*}
\end{center}

\subsection{Determining connections between servers}
\label{subsec:revealconnections}
The method to determine connections of a Bitcoin servers is similar to the
method of estimating the node's degree. It is based on sending marker
addresses to a peer which should then forward them to its
neighbours. The number of Bitcoin servers is estimated to be
about 10\% of the total number of Bitcoin peers. Note however that
clients can connect to the Bitcoin network only through connecting
to servers which means that all Bitcoin messages even those generated by
clients should travel along at least one link between two Bitcoin
servers.
We first describe a probabilistic method to determine
if two given peers \code{A} and \code{B} are connected which consists
of two phases.

During the first phase the attacker estimates the number
of connections of peer \code{A}. This number is used to compute the
number of marker addresses that will be forwarded to the peer's neighbours.
In the second phase the attacker chooses a set of fake marker addresses
$S=\{ip_1,ip_2,\dots,ip_{n}\}$ and sends them in \code{ADDR} messages to
peer \code{A} (10 addresses per message). She then sends \code{GETADDR}
messages to peer \code{B}. If the number of marker addresses known to \code{B}
corresponds to the number estimated in the first phase, node \code{B} is marked
as a neighbour.

An attacker can enhance this method to reveal a peer's connections by
applying it to each node in the list of running Bitcoin nodes (this
information is available to the attacker since running nodes advertise
their addresses). This is easily parallelized since the attacker needs to
send marker addresses just once. The drawbacks of the method is
that it does not allow to reveal connections to nodes which don't accept
incoming connection (e.g. located behind a NAT) since an attacker cannot
send \code{GETADDR} messages to such nodes.

Bitcoin network discovery protocol is designed in such a way so that
newly advertised addresses should be delivered to the majority of the nodes.
Thus one of the key ingredients of the method proposed in this section is
how to reduce the propagation radius. This is achieved by that each forwarded
address is accompanied with a timestamp. If this timestamp is older than
10 minutes, the address is not retransmitted anymore. Thus
in order to avoid false positives an attacker should set the timestamps
of the marker addresses to values close to 10 minutes to the past.

We implemented the method and carried out several experiments with our own
Bitcoin nodes which had 59, 53, 73, and 81 connections. As a list of candidates
we used 400 randomly chosen running Bitcoin nodes plus the nodes' current connections.
The results of the experiments are summarized in Table \ref{tab:numofconnections2}.
In all our experiments we had zero false positives.

In order to estimate how probable false positives are we chose 30 random
Bitcoin nodes and sent them marker addresses with timestamps set to 9 minutes 58 seconds
in the past relative the nodes' adjusted time. For each node we generated a unique set of
1000 marker addresses. At the same time we were monitoring
for received addresses at two of our nodes which had 83 and 85 connections.
At the end of the experiment no marker addresses arrived at our nodes which
indicated that false positives are quite unlikely.

\begin{center}
\begin{table}[h]
  \begin{tabular}{|p{1.6cm}|p{1.6cm}|p{1.6cm}|p{1.6cm}|}
    \hline
    Connections & Not behind NAT & Candidates & Discovered \\
    \hline
    59 & 25 & 459 & 25 \\
    \hline
    53 & 22 & 453 & 22 \\
    \hline
    73 & 8  & 473 & 8 \\
    \hline
    81 & 17 & 481 & 17 \\
    \hline
  \end{tabular}
  \label{tab:numofconnections2}
  \caption{Discovering Bitcoin node connections}
\end{table}
\end{center}

Finally in order to estimate the number of \code{GETADDR} messages sufficient to
learn addresses known to a peer we adopt a finite state discrete time Markov Chain
model~\cite{markovchains}.
Each state in the model represents  number of addresses learned by  the attacker.
At each step the attacker sends a \code{GETADDR} message and gets back 2500 random
addresses  from the total of maximum 20480 (note that some of those addresses may
already be known to the attacker from
the previous requests). The chain
has one absorbing state which is "all addresses are known to the attacker".
By computing fundamental matrix
we get the average number of transitions before the absorbing state is reached
which corresponds to the number of messages the attacker needs to send. If the
maximum number of addresses stored at node is 20480, it takes in average approximately
80 \code{GETADDR} messages to learn all those addresses.
Indeed, the probability for a single address to not be discovered is
upper bounded by $\left(\frac{2500 }{20480}\right)^{80}\approx \frac{1}{30000}$.
This estimation shows however an upper bound of the number of \code{GETADDR} messages.
Our experiments showed that it is sufficient to retrieve 5 \code{ADDR} message from
a peer in order to confirm that a connections exists, which significantly reduces
the number of \code{GETADDR} messages.

\section{CONCLUSION}

We have presented the first method that correlates pseudonyms of Bitcoin users behind NAT with the public IP address of the host where the transaction is generated. 
The crucial idea of our attack is to identify each client by an octet of outgoing connections it establishes. This octet of Bitcoin peers (entry nodes) serves as a unique identifier of a client
for the whole duration of a user session and will differentiate even those users who share the same NAT IP address. We showed that most of these connections can be learned if the attacker
maintains connections to a majority of Bitcoin servers. Then we show that the transaction propagation rules imply that the entry nodes will be among the first that report the transaction
to the attacker. As soon as the attacker receives the transaction from just 2-3 entry nodes he can with very high probability link the transaction to a specific client. Moreover a sequence of 
successfully mapped transactions can help the attacker to track dynamic changes in the entry node set, to keep the client identifier fresh.
The cost of the deanonymisation attack on the full Bitcoin network is under 1500 EUR.

We demonstrate that the use of Tor does not rule out the attack as Tor connections can be prohibited for the 
entire network. Our technique is orthogonal to the transaction graph de-anonymisation techniques and can be used in combination with them. 
It shows that the level of network anonymity provided by Bitcoin is quite low. Several features of the Bitcoin protocol makes the attack possible. 
In particular, we emphasize that the stable set of only 8 entry nodes is too small, as the majority of these nodes' connections can be captured by an attacker. A countermeasure could be to randomize and regularly rotate these nodes, and to submit transactions via another set of nodes.

We also described a number of techniques that reveal the topology of the Bitcoin network. Some of them are used for our attack, but the entire set is  interesting by themselves and not
only in the context of deanonymisation.
For example it can be used as a tool to better understand relations between
Bitcoin peers (e.g. one can check if peers of major pools keep permanent
connections between each other).
As another example, an adversary can find the
minimal cut in the network graph and target those connections with
denial-of-service attacks (an example of a memory exhaustion attack
that we discovered while digging through the Bitcoin source code can
be found in the appendix). This would result in splitting the network
in two parts. Our results open several directions for the future research. 

Yet another feature is the lack of authentication within the network,
which requires the nodes to blacklist misbehaving peers by IP. We figured
out that very short messages may cause a day IP ban, which can be used to
separate a given node or the entire network from anonymity
services such as proxy servers or Tor. If the Bitcoin community wishes
to use Tor, this part of the protocol must be reconsidered.

Finally, we showed that the routine procedure of adding a checkpoint
to the client code might be exploited to construct an alternate reality.
While too noticable as an attack scenario, this idea can be a solution
in the case of unforeseen and unsustainable rise of difficulty.

\bibliographystyle{plain}
\bibliography{bitcoin}
\appendix

\section{ESTIMATING SUCCESS RATE: DETAILS}
\label{sec:succ-det}

In this section we describe a mathematical model that allows us to estimate the success rate of the deanonymization attack.

As inputs, we take the average probability $p_{addr}$ over the network, which is estimated in Section~\ref{sec:an-num}, and the distribution of the number of entry nodes among the first 10 nodes
reporting a transaction to attacker's peers (Section~\ref{sec:experiments}). We extrapolate the latter probability spectrum from the test net to the main net, which assumes similar network performance and the stability of the spectrum when the attacker has more or fewer connections to servers. 
The correctness of the extrapolation can be tested only by mounting a full-scale attack on the network, which we chose not to perform for ethical reasons.

First, we introduce two combinatorial formulas. Suppose that there are $N$ balls. If each ball is red with probability  $p_a$, and green with probability $1-p_a$, then
 the probability that there are   $R$ red balls is
\begin{equation}\label{eq:p1}
\mathbb{P}_1(R;N) = \binom{N}{R}p_a^{j}(1-p_a)^{N-R}
\end{equation}

Now assume that there are $R$ red balls and $N-R$ green balls. Suppose that we select $L$ balls at random out of $N$. The probability that there will be exactly 
$q$ red balls among $L$ chosen is computed as follows:
$$
\mathbb{P}_2(q\,;\,L,R, N)  = \frac{\binom{R}{q}\binom{N-R}{L-q}}{\binom{N}{L}}.
$$  

Now we get back to Bitcoin. If each entry node is detected with probability $p_{addr}^{Avg} = 0.34$ (Section~\ref{sec:rate}), then according to Eq.~\eqref{eq:p1} we detect $R$ entry nodes out of 8 with the following probability
spectrum:
$$
\mathbb{P}_1(R;8):\quad \begin{array}{|c|c|}
\hline
\text{Nodes}&\text{Probability}\\
\hline
1 & 0.15\\
2& 0.27\\
3& 0.28\\
4& 0.18\\
5& 0.07\\
6&0.02\\
7&0.002\\
8&0.0002\\
\hline
\end{array}
$$
According to our experiments on  the Bitcoin test net (Section~\ref{sec:experiments}), the probability to have $L$ entry nodes among the top-10 is as follows:
$$
\mathbb{P}_3(L):\quad 
\begin{array}{|c|c|}
\hline
\text{Nodes}&\text{Probability}\\
\hline
1 & 0.02\\
2& 0.055\\
3& 0.1225\\
4& 0.245\\
5& 0.2125\\
6&0.2125\\
7&0.0925\\
8&0\\
\hline
\end{array}
$$

We assume that both events are independent. Then the probability that at least $M$ out of these $L$ nodes we have detected (i.e. it belongs to the set of $R$ entry nodes) is
$$
\mathbb{P}_{success}(M) = \sum_{q\geq M}\sum_{L\leq 8}\sum_{R\leq 8} \mathbb{P}_2(q\,;\,L,R, 8)\cdot \mathbb{P}_1(R;8)\cdot \mathbb{P}_3(L);
$$
We have made some calculations and got the following  results:
\begin{eqnarray*}
\sum_{L\leq 8}\sum_{R\leq 8} \mathbb{P}_2(q\,;\,L,R, 8)\cdot \mathbb{P}_1(R;8)\cdot \mathbb{P}_3(L):&\quad 
\begin{array}{|c|c|}
\hline
\text{L}&\text{Probability}\\
\hline
1 & 0.366\\
2& 0.243\\
3& 0.9\\
4& 0.02\\
5& 0.002\\
\hline
\end{array}\\[5pt]
\mathbb{P}_{success}(M):&\quad 
\begin{array}{|c|c|}
\hline
M&\text{Probability}\\
\hline
1 & 0.721\\
2& 0.355\\
3& 0.112\\
4& 0.022\\
5& 0.002\\
\hline
\end{array}
\end{eqnarray*}
Therefore, we expect to catch 3-tuples in 11\% of transactions, and 2-tuples in 35\% of transactions.

We applied this model to the testnet as well, and obtained that it fits our actual deanonymization results well:
$$
\begin{array}{|c|c|c|}
\hline
\text{Estimated }p_{addr} &\multicolumn{2}{c|} {\text{Deanonymization rate with 3-tuples}}\\
\cline{2-3}
& \text{Actual} & \text{Predicted}\\
\hline
0.64 &  41\% & 43\%\\
0.86 & 59.9\%& 65.6\%\\
\hline
\end{array}
$$

\section{ATTACK COSTS}

\label{sec:costs}

The expenses for the attack include two main components: (1) renting machines for connecting to
Bitcoin servers and listening for \code{INVENTORY} messages; (2) periodically advertising
potential client addresses in the network. Note that if an attacker rents servers,
the incoming traffic for the servers is normally free of charge.
Assuming that an attacker would like to stay stealthy, she would want to
have 50 different IP addresses possibly from different subnetworks. Thus
she might want to rent 50 different servers. Assuming monthly price per
one server 25 EUR, this results in 1250 EUR per month.

When advertising potential client addresses, the attacker is interested in that the
addresses propagate in the network as fast as possible. In order to achieve this
the attacker might try to advertise the addresses to all servers simultaneously.
Given that there are 100,000 potential clients and the attacker needs to send 
10 addresses per \code{ADDR} message, this results in 10,000 \code{ADDR} messages
of 325 bytes each per Bitcoin server or (given there are 
8,000 Bitcoin servers) 24.2 GB in total.

If an attacker advertises the addresses every 10 minutes and she is interested
in continuously deanonymising transaction during a month, it will require sending
104,544 GB of data from 50 servers. Given that 10,000 GB per server is included
into the servers price and the price per additional 1,000 GB is 2 EUR, 
the attacker would need to pay 109 EUR per month.
As a result the total cost of the attack is estimated to be less than 1500 EUR per month
of continuous deanonymisation.

\section{TRANSACTION PROPAGATION DELAY}
\begin{figure}[h]
\begin{center}
\includegraphics[scale=0.4]{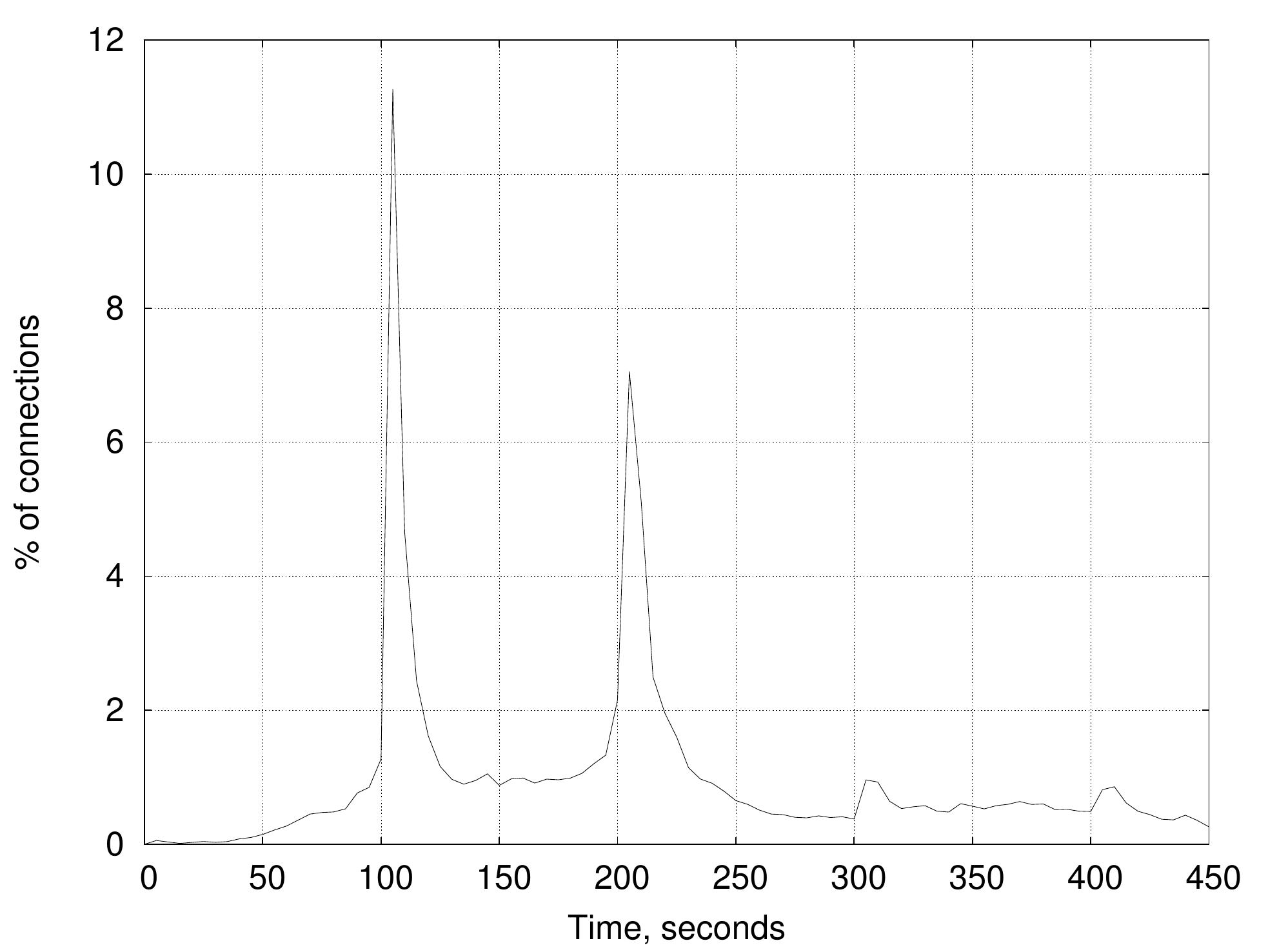}
\caption{Transaction propagation delay, density}
\label{fig:delays-density}
\end{center}
\end{figure}

\begin{figure}[h]
\begin{center}
\includegraphics[scale=0.4]{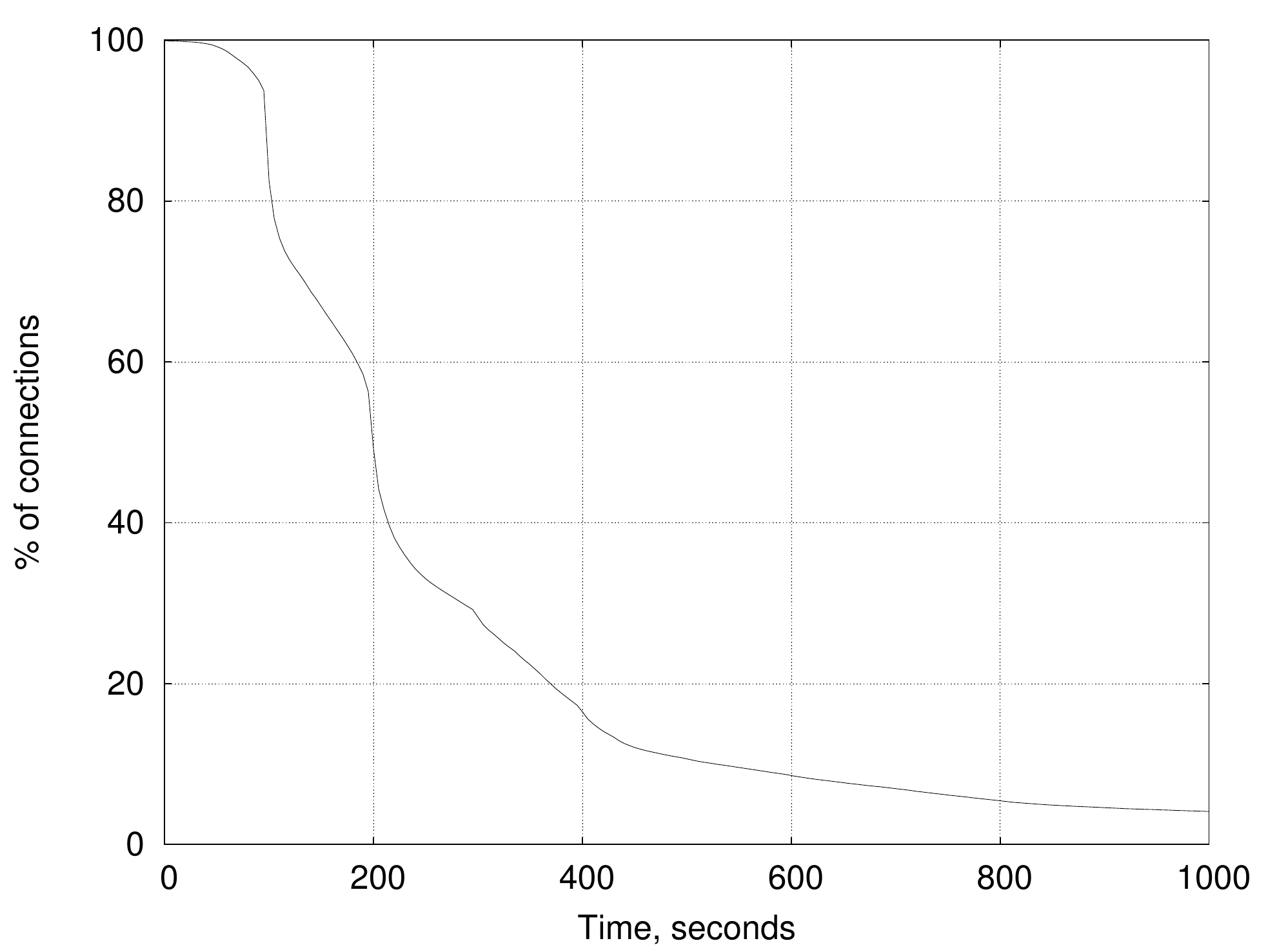}
\caption{Transaction propagation delay, cumulative}
\label{fig:delays-cumulative}
\end{center}
\end{figure}

In this section we measure transaction propagation delays between
our high-speed server (1 Gbit/s, Intel Core i7 3GHz) 
located in Germany and 6,163 other Bitcoin servers.
As was described in Section~\ref{sec:background}, it takes 3 steps
to forward a transaction between two Bitcoin peers.
As we are not able to obtain times when a remote peer sends
an \code{INVENTORY} message, we skipped the first step (i.e. propagation delays
of \code{INVENTORY} messages)
and measured time differences between receptions
of corresponding \code{INVENTORY} messages and receptions of the
transactions. Note however that the size of an \code{INVENTORY} message
is 37 bytes, while the size of a transaction which transfers coins from
one pseudonym to two other pseudonyms is 258 bytes. Thus
the obtained results can serve as a good approximation.
For each Bitcoin server we collected 70 transactions and combined them
into a single dataset (thus having 431,410 data points).
Fig.~\ref{fig:delays-density} shows probability density function of the
transaction propagation delay between our node and other Bitcoin servers and
Fig.~\ref{fig:delays-cumulative} shows the corresponding cumulative distribution.

\section{ON STABILITY OF THE FINGERPRINT}
\label{sec:entry-stability}
In this section we estimate the stability of a client's fingerprint (the
set of eight first-hop connections). According to the \textsf{bitcoind}
source code (version v0.9.1), there are three reasons why an entry node
can be disconnected from a client:
\begin{itemize}
  \item The client switched off the computer/closed Bitcoin application.
  \item No data was sent over a connection for 1.5 hours.
  \item An Entry node goes offline.
\end{itemize}
Given the number of transitions generated by the network\cite{blockchain},
block generation rate, and addresses propagation,
some data is normally sent to and from the entry nodes within 1.5 hours.

\begin{figure}[h]
\begin{center}
\includegraphics[scale=0.6]{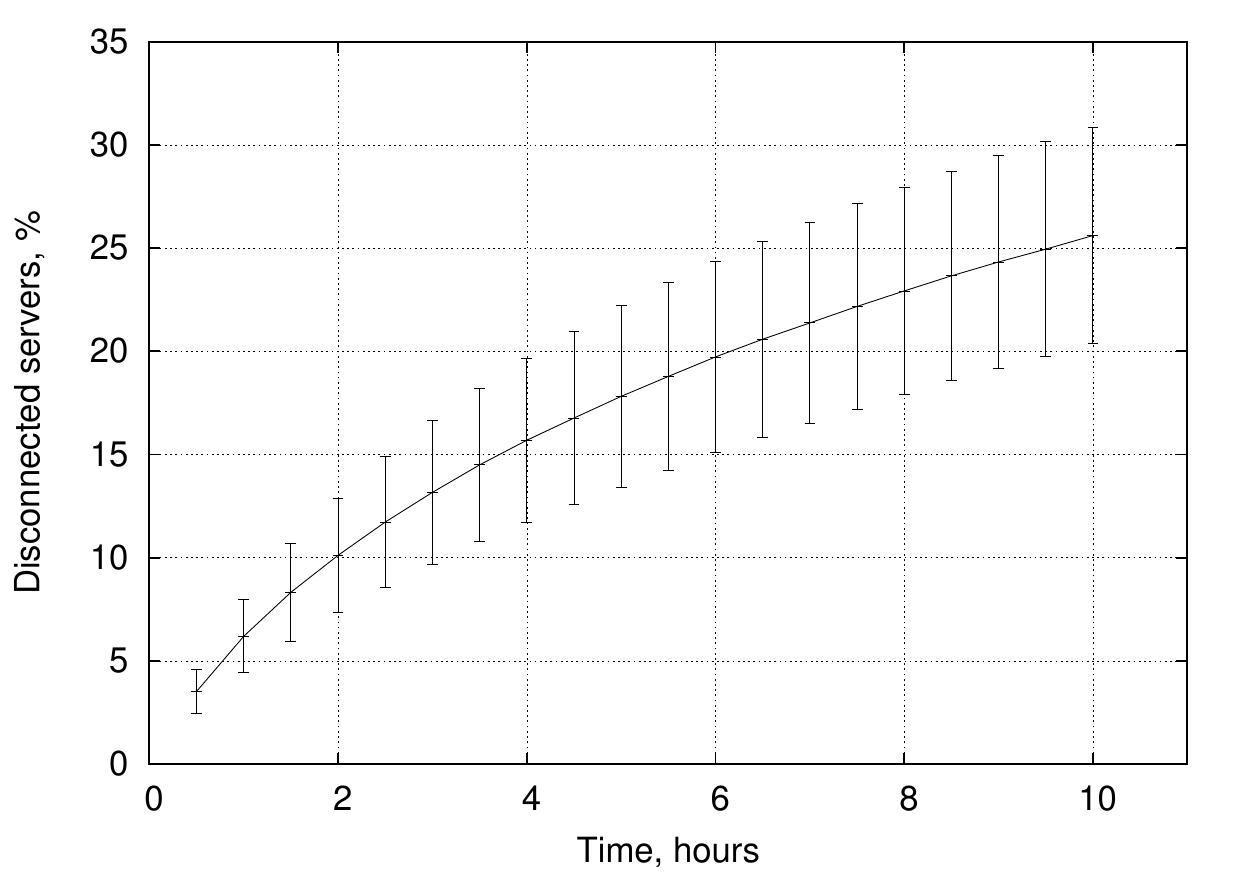}
\caption{Bitcoin servers churn rate}
\label{fig:servers-mainnet-disconnect-rate}
\end{center}
\end{figure}

In order to estimate the probability of an entry node going off-line we
we took data from http://getaddr.bitnodes.io which produces a list of
running Bitcoin servers every five minutes. We analysed the data for two
weeks. The probability for a node to disconnect after specific amount
of time with 95\% confidence interval is shown on
Fig.~\ref{fig:servers-mainnet-disconnect-rate}.

Fig.~\ref{fig:servers-mainnet-disconnect-rate} shows that after 2.5 hours
only one node would disconnect on average and only two nodes will disconnect
after 10 hours. So for the typical duration of a client session the
fingerprint is very stable.
In our experiment, after running our Bitcoin client for about 10 hours 3 nodes
out of eight have disconnected.

The second point we address in this section is regarding the usage of VPN
which is a popular recommendation for preserving anonymity in Bitcoin~\cite{bitcoinvpn}.
While protecting a user's IP, the stability of the fingerprint still allows
an attacker to glue together different Bitcoin addresses of the same user.
We checked the stability of the fingerprint on the Bitcoin testnet while
connecting to the network:
\begin{enumerate}
  \item via public free VPNs (vpngate.net);
  \item via a non-free one (AirVPN).
  \item via our own VPN server.
\end{enumerate}
For cases 2 and 3, the stability of the fingerprint was the same as if no
VPN was used.
For case 1, connections to entry nodes were dropped
from time to time (about every 20 mins for the main net and about every
few minutes for the testnet due too absence of traffic) by the VPN servers.
It's likely that free VPN servers were set with small inactivity timeouts
and some limits for connection durations.

\section{DENIAL OF SERVICE}
\label{sec:dosing}
In this section we analyse the security of Bitcoin networking
protocol against Denial of Services attacks.

\subsection{Memory exhaustion by address flooding}
\label{subsec:memexhaust}
Bitcoin's peer discovery protocol has a mechanism which prevents multiple
retransmissions of the same addresses: for each connection it has, a
Bitcoin node maintains a history (Implemented as an instance of
std::set C++ class) of addresses which were sent over this connection.
This history is emptied once per every 24 hours and more importantly does
not limit the number of elements it holds. In order to check if one can
flood this container with fake addresses we conducted a simplified
experiment. We set up locally two Bitcoin nodes so that when one of the
nodes (the target) receives an \code{ADDR} message it forwards the addresses
it contains to just one neighbour. Both machines had Ubuntu 12.04 installed
with 2Gb of RAM and the same amount of swap memory. They were running
\textsf{bitcoind} version 0.8.6.

We were sending fake addresses with the rate of 30,000 addresses per second.
After approximately 45 minutes, the response delay to the user's interactions
became significant and the node was unreachable for new Bitcoin connections.
We also mounted a reduced version of this attack on our own Bitcoin node
in the real network.  We terminated the experiment when the memory consumption
increased by 100 MB.


\end{document}